\documentclass[conference]{IEEEtran}
\IEEEoverridecommandlockouts
\AtBeginDocument{%
  \providecommand\BibTeX{{%
    \normalfont B\kern-0.5em{\scshape i\kern-0.25em b}\kern-0.8em\TeX}}}

\usepackage{algorithm}
\usepackage[noend]{algorithmic}
\usepackage{float}
\usepackage{caption}
\usepackage{lipsum}
\usepackage{xcolor}
\usepackage{tabularx}
\usepackage{placeins}
\usepackage{enumitem}
\usepackage{booktabs}
\usepackage{multirow}
\usepackage{array}
\usepackage{dblfloatfix}
\usepackage{amsmath}
\usepackage{amssymb}
\usepackage{tikz}
\usepackage{placeins}
\usetikzlibrary{matrix, fit, positioning, backgrounds, shapes.geometric, calc, arrows.meta, plotmarks}
\usepackage{pgfplots}
\usepgfplotslibrary{groupplots}
\pgfplotsset{compat=1.17}
\usepackage{cite}
\usepackage{hyperref}
\hypersetup{colorlinks=true,urlcolor=blue,linkcolor=black,citecolor=black}
\usepackage{listings}
\lstdefinestyle{sqllist}{
  basicstyle=\ttfamily\footnotesize,
  breaklines=true,
  frame=single,
  framesep=4pt,
  xleftmargin=1em,
  aboveskip=6pt,
  belowskip=6pt,
  showstringspaces=false,
  columns=fullflexible,
  keepspaces=true
}

\newtheorem{myEmp}{\bf Example}

\newtheorem{myProblem}{\bf Problem}

\usepackage[tikz]{bclogo}
\usetikzlibrary{positioning,arrows.meta,fit,calc}

\lstnewenvironment{sqlblock}{\lstset{style=sqllist}}{}

\providecommand{\spara}[1]{\smallskip\noindent{\bf #1}}

\begin{document}

\title{Efficient Semantic ORDER BY via Transitivity-Aware LLM Batch Sorting}

\author{\IEEEauthorblockN{Akinkunle Akintayo Akinola \,\, Arijit Khan}
\IEEEauthorblockA{
\textit{Bowling Green State University, USA}\\
akinola@bgsu.edu \,\, arijitk@bgsu.edu}
}

\maketitle

\begin{abstract}
Large language models (LLMs) enable declarative semantic sorting through \texttt{LLM-ORDER BY} operators that rank tuples according to natural-language criteria over unstructured data. Executing such operators efficiently is challenging because LLM comparisons are expensive, stochastic, and sometimes inconsistent. Pairwise approaches require many costly calls. Listwise approaches such as external merge sort rank several items per call, but they follow a fixed merge schedule and keep no record of earlier answers: each merge call re-asks orderings that previous calls have already revealed, directly or by transitivity. We present \emph{Transitive Batched Sort} (\textsf{TBS}), a transitivity-aware physical operator for semantic sorting that records every revealed preference in a global acyclic preference graph and issues a listwise request only for items whose relative order is not yet implied. To scale, \textsf{TBS} stores only directly observed preferences and answers transitivity queries through a memory-efficient reachability index; to handle noisy LLM outputs, it validates responses and rejects cycle-inducing updates. We evaluate TBS on six benchmarks spanning structured sorting, product price ranking, passage ranking, and personalized recommendation with four LLMs. \textsf{TBS} reduces API calls by $12$--$20\times$ relative to pairwise quicksort and, at the same batch size, uses 1.1–1.6× fewer calls and $1.3$--$5.3\times$ lower cost than listwise merge sort while achieving competitive or higher ranking quality across most benchmarks; on {\em Amazon} it scales to one million products. These results show that reusing previously inferred preference relations, rather than re-querying them, is an effective foundation for scalable semantic \texttt{ORDER BY} operators.
\end{abstract}


\section{Introduction}
\label{sec:intro}

\spara{Declarative Semantic Sorting in Database Engines.}
Integrating Large Language Models (LLMs) into relational query engines enables declarative querying over unstructured attributes using subjective, natural-language criteria~\cite{patel2024semantic,zhao2025accesspaths}.
Unlike supervised learning-to-rank models~\cite{liu2009learningtorank}, which require labeled training data, or retrieval pipelines built on pre-computed embedding indices~\cite{karpukhin2020dense,johnson2019billion}, an LLM can interpret a new criterion zero-shot at query time.
Recent systems expose this capability as semantic operators that extend SQL-style dataflows, including a declarative semantic sorting clause, \texttt{LLM-ORDER BY}, that ranks tuples by a natural-language criterion~\cite{patel2024semantic,zhao2025accesspaths}.
While declarative specification simplifies query authoring, executing semantic sorting efficiently is difficult.
Classical sort operators~\cite{graefe2006implementing} assume cheap, deterministic, and transitive comparisons, whereas LLM comparators are expensive, stochastic, and can be inconsistent.
Existing \texttt{LLM-ORDER BY} operators~\cite{patel2024semantic, zhao2025accesspaths,listk2026} reduce cost through access-path selection or pivot-based schedules, but they lack \emph{physical operators} that retain and reuse the preferences revealed by earlier LLM calls.

\begin{myEmp}[Situation report prioritization with implicit urgency]
\label{ex:sitrep}
During emergency disaster response, an operations center maintains real-time situation reports in a relational table: \texttt{SITREP (id, ward, category, reported\_at, description)}, containing hundreds of streaming incident logs.
The \texttt{description} attribute contains unstructured text, e.g., \emph{``Chest-deep water blocking access road to Riverside Nursing Home''}, or \emph{``Minor pooling in parking lot; no injuries.''}

\smallskip
\noindent\textbf{Explicit \textsc{order by}.}
An analyst attempting to prioritize severe incidents using standard SQL must engineer brittle, hand-tuned scoring heuristics:
\begin{sqlblock}
SELECT id FROM SITREP
ORDER BY 0.15 * recency_score(reported_at)
       + 0.85 * keyword_hits(description, CRITICAL_TERMS) DESC;
\end{sqlblock}
Such explicit scoring functions may fail to capture nuanced domain semantics and require continuous manual retuning as operational conditions change.
Even enhancing such heuristics with supervised learning-to-rank models~\cite{liu2009learningtorank} requires labeled training data and task-specific feature engineering that are impractical in rapidly evolving operational workloads.

\smallskip
\noindent\textbf{Semantic \textsc{order by}.}
In contrast, declarative semantic sorting delegates domain understanding to an LLM via natural language:
\begin{sqlblock}
SELECT id FROM SITREP
LLM-ORDER BY "Rank by immediate response urgency, prioritizing life safety,
              critical-facility access, and vulnerable populations.";
\end{sqlblock}
\end{myEmp}

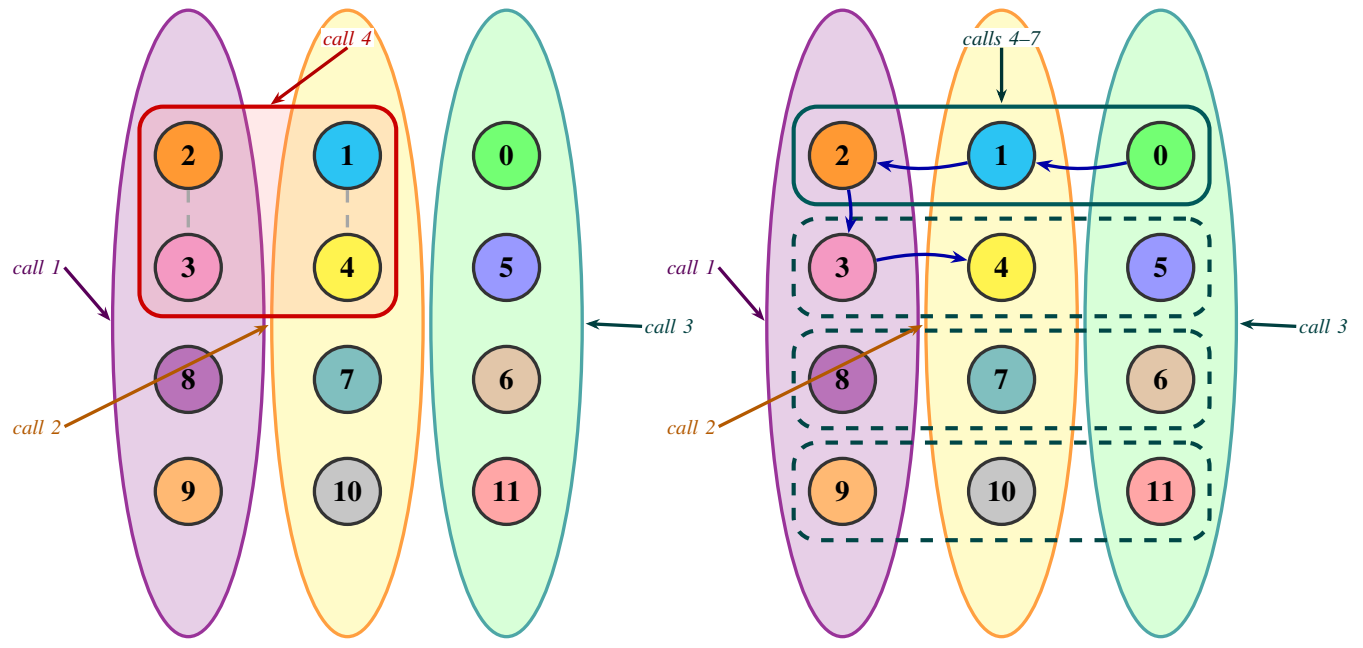
\begin{figure}[t]
\centering
\resizebox{\columnwidth}{!}{%
\begin{tikzpicture}[
 font=\scriptsize,
 key/.style={circle,draw=black!80,thick,minimum size=5.6mm,font=\scriptsize\bfseries,inner sep=0pt},
 blob/.style={ellipse,inner sep=1.6mm,fill opacity=.34,line width=.8pt,outer sep=0pt},
 rowbox/.style={rectangle,rounded corners=2mm,inner sep=1.2mm,draw,line width=1.0pt,fill=none,outer sep=0pt},
 callab/.style={font=\tiny\itshape,inner sep=0.5pt,align=center,fill=white,draw=none},
 arr/.style={-{Stealth[length=1.2mm]},thick,shorten >=0.8pt}
]
% -------- (a) Listwise merge sort --------
\begin{scope}
\node[key,fill=orange!80] (a2) at (0.00,2.85) {2};
\node[key,fill=cyan!65]   (a1) at (1.35,2.85) {1};
\node[key,fill=green!55]  (a0) at (2.70,2.85) {0};
\node[key,fill=magenta!50](a3) at (0.00,1.90) {3};
\node[key,fill=yellow!80] (a4) at (1.35,1.90) {4};
\node[key,fill=blue!40]   (a5) at (2.70,1.90) {5};
\node[key,fill=violet!55] (a8) at (0.00,0.95) {8};
\node[key,fill=teal!50]   (a7) at (1.35,0.95) {7};
\node[key,fill=brown!45]  (a6) at (2.70,0.95) {6};
\node[key,fill=orange!55] (a9) at (0.00,0.00) {9};
\node[key,fill=gray!45]   (a10) at (1.35,0.00) {10};
\node[key,fill=red!35]    (a11) at (2.70,0.00) {11};
\begin{scope}[on background layer]
\node[blob,fill=violet!55,draw=violet!80,fit=(a2)(a3)(a8)(a9)] (ac0) {};
\node[blob,fill=yellow!75,draw=orange!75,fit=(a1)(a4)(a7)(a10)] (ac1) {};
\node[blob,fill=green!45,draw=teal!70,fit=(a0)(a5)(a6)(a11)] (ac2) {};
\node[rowbox,draw=red!80!black,fill=red!55,fill opacity=.16,fit=(a2)(a1)(a3)(a4)] (amerge) {};
\end{scope}
\draw[gray!70,dashed,thick] (a2)--(a3);
\draw[gray!70,dashed,thick] (a1)--(a4);
\node[callab,text=violet!70!black,anchor=east] (al1) at (-1.05,1.90) {call 1};
\node[callab,text=orange!70!black,anchor=east] (al2) at (-1.05,0.55) {call 2};
\node[callab,text=teal!50!black,anchor=west] (al3) at (3.85,1.40) {call 3};
\node[callab,text=red!70!black] (al4) at (1.35,3.85) {call 4};
\draw[arr,violet!70!black] (al1.east)--(ac0.west);
\draw[arr,orange!70!black] (al2.east)--(ac1.west);
\draw[arr,teal!50!black] (al3.west)--(ac2.east);
\draw[arr,red!70!black] (al4.south)--(amerge.north);
\node[font=\scriptsize,anchor=north] at (1.35,-1.55)
  {\textbf{(a)} Listwise merge sort (11 calls)};
\end{scope}
% -------- (b) TBS --------
\begin{scope}[shift={(5.55,0)}]
\node[key,fill=orange!80] (b2) at (0.00,2.85) {2};
\node[key,fill=cyan!65]   (b1) at (1.35,2.85) {1};
\node[key,fill=green!55]  (b0) at (2.70,2.85) {0};
\node[key,fill=magenta!50](b3) at (0.00,1.90) {3};
\node[key,fill=yellow!80] (b4) at (1.35,1.90) {4};
\node[key,fill=blue!40]   (b5) at (2.70,1.90) {5};
\node[key,fill=violet!55] (b8) at (0.00,0.95) {8};
\node[key,fill=teal!50]   (b7) at (1.35,0.95) {7};
\node[key,fill=brown!45]  (b6) at (2.70,0.95) {6};
\node[key,fill=orange!55] (b9) at (0.00,0.00) {9};
\node[key,fill=gray!45]   (b10) at (1.35,0.00) {10};
\node[key,fill=red!35]    (b11) at (2.70,0.00) {11};
\begin{scope}[on background layer]
\node[blob,fill=violet!55,draw=violet!80,fit=(b2)(b3)(b8)(b9)] (bc0) {};
\node[blob,fill=yellow!75,draw=orange!75,fit=(b1)(b4)(b7)(b10)] (bc1) {};
\node[blob,fill=green!45,draw=teal!70,fit=(b0)(b5)(b6)(b11)] (bc2) {};
\node[rowbox,draw=teal!70!black,fit=(b2)(b1)(b0)] (br0) {};
\node[rowbox,draw=teal!55!black,dashed,fit=(b3)(b4)(b5)] (br1) {};
\node[rowbox,draw=teal!55!black,dashed,fit=(b8)(b7)(b6)] (br2) {};
\node[rowbox,draw=teal!55!black,dashed,fit=(b9)(b10)(b11)] (br3) {};
\end{scope}
\draw[-{Stealth[length=1.4mm]},thick,blue!65!black] (b0) to[bend left=12] (b1);
\draw[-{Stealth[length=1.4mm]},thick,blue!65!black] (b1) to[bend left=12] (b2);
\draw[-{Stealth[length=1.4mm]},thick,blue!65!black] (b2) to[bend left=10] (b3);
\draw[-{Stealth[length=1.4mm]},thick,blue!65!black] (b3) to[bend left=12] (b4);
\node[callab,text=violet!70!black,anchor=east] (bl1) at (-1.05,1.90) {call 1};
\node[callab,text=orange!70!black,anchor=east] (bl2) at (-1.05,0.55) {call 2};
\node[callab,text=teal!50!black,anchor=west] (bl3) at (3.85,1.40) {call 3};
\node[callab,text=teal!40!black] (bl4) at (1.35,3.85) {calls 4--7};
\draw[arr,violet!70!black] (bl1.east)--(bc0.west);
\draw[arr,orange!70!black] (bl2.east)--(bc1.west);
\draw[arr,teal!50!black] (bl3.west)--(bc2.east);
\draw[arr,teal!40!black] (bl4.south)--(br0.north);
\node[font=\scriptsize,anchor=north,align=center] at (1.35,-1.55)
  {\textbf{(b)} \textsf{TBS} (7 calls)};
\end{scope}
\end{tikzpicture}%
}
\caption{\small Walk-through with $m{=}4$ on shuffled input
$X{=}(8,9,3,2,10,4,7,1,0,11,6,5)$
(ascending gold order $0{\prec}1{\prec}\cdots{\prec}11$).
The drawing is the $4{\times}3$ grid \emph{after} Round~1.
\textbf{Shared Round~1 (3 calls).}
Call~1 sorts $\{8,9,3,2\}$ to $2{\prec}3{\prec}8{\prec}9$;
call~2 sorts $\{10,4,7,1\}$ to $1{\prec}4{\prec}7{\prec}10$;
call~3 sorts $\{0,11,6,5\}$ to $0{\prec}5{\prec}6{\prec}11$.
\textbf{(a)~Listwise merge sort}~\cite{zhao2025accesspaths} (Alg.~4)
then merges the three sorted runs with $h{=}\lfloor m/2\rfloor{=}2$
(merge batch size up to~$m$).
Call~4 ranks $\{2,3,1,4\}$ (red box) as $1{\prec}2{\prec}3{\prec}4$.
The dashed pairs $2{\prec}3$ and $1{\prec}4$ were already returned by
calls~1--2: the size-$4$ merge \emph{re-asks} them.
Seven further merge calls follow, each likewise repeating at least one
already-known in-run pair, for \textbf{11} LLM calls total.
\textbf{(b)~\textsf{TBS}} stores the same Round~1 answers in a DAG.
Round~2 here sorts every row, as in Algorithm~\ref{alg:schedule} (the experiments' default sorts only $R_0$, \S\ref{sec:method:optimizations}).
Round~2 sorts the four rows (calls~4--7):
$\{2,1,0\}{\to}0{\prec}1{\prec}2$,
$\{3,4,5\}{\to}3{\prec}4{\prec}5$,
$\{8,7,6\}{\to}6{\prec}7{\prec}8$,
$\{9,10,11\}{\to}9{\prec}10{\prec}11$
(solid box ${=}$ call~4; dashed boxes ${=}$ calls~5--7).
Transitivity closes the full chain
$0{\prec}1{\prec}\cdots{\prec}11$, so $u{=}0$:
no diagonal calls and no packing.
\textbf{Total: 7 LLM calls} (vs.\ 11 for merge).}
\label{fig:intro-transitivity}
\vspace{-5mm}
\end{figure}

\spara{The Physical Execution Bottleneck.}
While conceptually straightforward, executing \texttt{LLM-ORDER BY} inside a relational query engine violates classical database sorting assumptions~\cite{graefe2006implementing}, which rely on cheap, deterministic, and strictly transitive pairwise comparators to achieve $O(n \log n)$ complexity.
Generative LLM comparators violate every one of these assumptions: they are expensive (each call incurs monetary API cost and network latency), stochastic (non-deterministic across invocations), and non-transitive (produce cyclic preferences).
Existing LLM-augmented sorting approaches~\cite{zhao2025accesspaths, dai2023uncovering} exhibit critical limitations:

\begin{itemize}[leftmargin=*]
    \item \textbf{Costly Pairwise Comparisons:} 
    Classic sorting algorithms such as quicksort and merge sort rely on pairwise comparisons \cite{dai2023uncovering,qin2024large}. This is acceptable for classical sorting operators \cite{graefe2006implementing}, where a comparison is nearly free, but with LLMs each comparison is a costly API call that settles a single pair, even though one call can order a larger batch. Comparator cost therefore dominates query latency (e.g., about 1,500 calls for Quicksort at n = 200).
    \item \textbf{Pointwise Calibration:} Pointwise scoring assigns an independent scalar rating to each tuple~\cite{dai2023uncovering}.
    It is cheap and accurate when every key has an absolute attribute the model can recall, such as a player's height.
    Scores from separate prompts are not calibrated against each other, however, so pointwise degrades on relative criteria such as relevance or preference (\S\ref{sec:exp:rq1}).
    \item \textbf{Redundant Listwise Merge Calls:}
    Listwise prompting amortizes cost by ranking batches of size~$m$~\cite{tang2024permutation,ma2023listwise,zhao2025accesspaths}.
    The strongest published listwise \texttt{LLM-ORDER BY} schedule is OrderByLLM external merge sort~\cite{zhao2025accesspaths}: it sorts contiguous runs of~$m$ keys, then merges those runs with further listwise calls of size up to~$m$ (Alg.~4; merge width $h{=}\lfloor m/2\rfloor$ heads from each run).
    That schedule never stores the pairwise preferences a call reveals, so a merge batch can \emph{re-ask} in-run pairs that Round~1 already ordered---even when the merge prompt itself uses the full batch budget~$m$.
    Figure~\ref{fig:intro-transitivity} makes this concrete for $m{=}4$ on $n{=}12$: after three shared Round~1 calls, merge call~4 ranks $\{2,3,1,4\}$ and re-queries $2{\prec}3$ and $1{\prec}4$, which calls~1--2 already returned; seven further merge calls each repeat at least one known in-run pair (11 calls total).
    Sliding-window rerankers have an analogous issue when overlapping windows re-rank the same keys.
    Neither family can detect that a later batch is already implied by earlier answers.
    \item \textbf{Stochastic Noise and Cycle Inconsistencies:} LLM outputs frequently exhibit positional bias, missing identifiers, duplicates, and non-transitive preference cycles ($A \succ B \succ C \succ A$). A practical database engine must enforce strict partial-order consistency over noisy, malformed generative replies.
\end{itemize}

\spara{Our Approach: Transitive Batched Sort (\textsf{TBS}).}
To address these challenges, we propose \textsf{TBS}, a transitivity-aware
physical operator for semantic sorting.
Unlike classical database sort operators~\cite{graefe2006implementing}, which
assume cheap and deterministic comparisons, and unlike OrderByLLM merge
sort~\cite{zhao2025accesspaths}, which discards preferences after each call,
\textsf{TBS} is built around expensive listwise LLM calls \emph{and} a persistent
preference DAG.
It organizes keys into a fixed three-round batch schedule (column sorting,
recursive row sorting, and diagonal stitching) whose rounds compare largely
disjoint sets of pairs, and records every accepted preference as a DAG edge.
Before issuing a scheduled batch, \textsf{TBS} tests reachability and skips the
call if every pair in the batch is already implied---the reason the $m{=}4$
walk-through in Figure~\ref{fig:intro-transitivity} finishes in $7$ calls
instead of merge's~$11$, with Round~2 rows closing the total order by
transitivity and no diagonal or packing calls required.
Rather than materializing the quadratic transitive closure, \textsf{TBS} stores
only direct edges and answers reachability through a dynamic topological-order
index (linear memory in keys and stored preferences).
To handle noisy LLM outputs, it validates each reply with six checks and rejects
cycle-inducing edges; to control cost, it stops early once a target fraction of
pairs is ordered and selects~$m$ automatically.

\spara{Our Contributions.}
We make the following contributions:
\begin{itemize}[leftmargin=*]
    \item \textbf{Transitivity-Aware Physical Operator:} We formalize declarative \texttt{LLM-ORDER BY} and design \textsf{TBS}, a three-round batched schedule that records every listwise answer in a preference graph and skips batches whose pairs are already implied by transitivity, eliminating the re-ask of known pairs that listwise merge sort incurs even at full merge batch size~$m$ (Figure~\ref{fig:intro-transitivity}).
    Unlike classical $k$-sorter and grid schedules~\cite{beigel1990ksorter,leighton1985columnsort,scherson1986shearsort}, which assume a perfect comparator, \textsf{TBS} skips batches already settled by earlier answers and rejects contradictions.
    Cycle-inducing updates are rejected, so the graph stays acyclic and \textsf{TBS} always returns a total order consistent with all accepted preferences.
    \item \textbf{Memory-Efficient Reachability Index:} We replace explicit transitive closure with a direct-edge DAG and a dynamic topological-order index. The index makes the same decisions as explicit closure while using linear rather than quadratic memory, and in our experiments it is $45$--$1{,}200\times$ faster with $25$--$83\times$ less peak memory (median cache-replay sort time after warm-up; peak Python heap via \texttt{tracemalloc}).
    \item \textbf{Robustness, Tuning for Noisy LLMs, and Optimization:} We introduce a six-check validation-and-repair layer, cycle-inducing update rejection, completion-based early stopping, and knee-based batch-size selection. We also study how Round~1 membership presets affect quality and cost.
    \item \textbf{Comprehensive Evaluation:} On six benchmarks and four LLMs, \textsf{TBS} uses $12$--$20\times$ fewer API calls than pairwise Quicksort~\cite{zhao2025accesspaths} and fewer calls than OrderByLLM listwise merge sort at matched~$m$ on every dataset (e.g., $97$ vs.\ $148$ on {\em NBA} and $1{,}790$ vs.\ $2{,}347$ on {\em DL19} with GPT-4o), while staying competitive with merge quality and leading mean full-list nDCG on {\em DL19} and {\em SciFact} (within $0.008$ of QS$_1$ on {\em MovieLens}).
\end{itemize}

\section{Related Work}
\label{sec:related}

This work lies within the emerging area of AI-augmented database systems, where LLMs are integrated into query execution to support semantic operators over unstructured data~\cite{patel2024semantic,zhao2025accesspaths,listk2026}. We focus on semantic \texttt{LLM-ORDER BY}, which aims to produce a global ranking of a supplied key list according to a natural language criterion. Prior work differs primarily in the interaction paradigm used to obtain rankings from an LLM, including pointwise, pairwise, and listwise approaches. We review these paradigms and discuss their implications for semantic sorting.

\begin{table}[tb!]
  \centering
  \caption{\small Comparison with selected prior work on LLM-based ranking and
  semantic \textsc{ORDER BY}. Columns are grouped into \emph{scope},
  \emph{state reuse}, and \emph{robustness and analysis}; $\checkmark$ means
  the property is present, -- means it is absent or not addressed.  \textbf{Op}: targets a database semantic \textsc{ORDER BY} over a supplied key list. 
  \textbf{ZS}: uses zero-shot general-purpose chat models. 
  \textbf{Full}: materializes a global total order. 
  \textbf{POG}: maintains an explicit global graph of accepted preferences across calls. 
  \textbf{Transitivity}: queries that graph for transitively implied relations and skips a call when all its pairs are already implied. 
  \textbf{Validation}: repairs malformed outputs. 
  \textbf{Batch-size}: tunes listwise batch size. 
  \textbf{Membership}: ablates batch record membership and prompt order. 
  \textbf{Errors}: formalizes sorting errors separately from format repair.}
  \label{tab:related-compare}
  \scriptsize
  \setlength{\tabcolsep}{2.5pt}
  \renewcommand{\arraystretch}{1.12}
  % requires \usepackage{graphicx}
  \newcommand{\rh}[1]{\rotatebox[origin=c]{90}{\textbf{#1}}}
  \begin{tabular}{@{}l*{9}{c}@{}}
    \toprule
    & \multicolumn{3}{c}{\textbf{Scope}}
    & \multicolumn{2}{c}{\textbf{State reuse}}
    & \multicolumn{4}{c}{\textbf{Robustness / analysis}} \\
    \cmidrule(lr){2-4} \cmidrule(lr){5-6} \cmidrule(lr){7-10}
    \textbf{Method} & \rh{Op} & \rh{ZS} & \rh{Full}
      & \rh{POG} & \rh{Transitivity}
      & \rh{Validation} & \rh{Batch-size} & \rh{Membership} & \rh{Errors} \\
    \midrule
    OrderByLLM~\cite{zhao2025accesspaths} & $\checkmark$ & $\checkmark$ & $\checkmark$ & -- & -- & $\checkmark$ & -- & -- & -- \\
    LOTUS~\cite{patel2024semantic} & $\checkmark$ & $\checkmark$ & $\checkmark$ & -- & -- & -- & -- & -- & -- \\
    ListK~\cite{listk2026} & $\checkmark$ & -- & $\checkmark$ & -- & -- & -- & -- & -- & -- \\
    LLM-RankFusion~\cite{zeng2024rankfusion} & -- & $\checkmark$ & -- & -- & -- & -- & -- & -- & -- \\
    LLM rerankers~\cite{sun2023chatgpt,ma2023listwise,qin2024large} & -- & $\checkmark$ & -- & -- & -- & -- & -- & -- & -- \\
    Setwise~\cite{zhuang2024setwise} & -- & $\checkmark$ & -- & -- & -- & -- & -- & -- & -- \\
    Trained LTR rerankers~\cite{nogueira2019passage,liu2009learningtorank} & -- & -- & -- & -- & -- & -- & -- & -- & -- \\
    \midrule
    \textbf{\textsf{TBS} (Ours)} & $\checkmark$ & $\checkmark$ & $\checkmark$ & $\checkmark$ & $\checkmark$ & $\checkmark$ & $\checkmark$ & $\checkmark$ & $\checkmark$ \\
    \bottomrule
  \end{tabular}
\end{table}

\subsection{Pointwise, Pairwise, and Listwise Paradigms}
\label{sec:related:paradigms}

Classical database sorting assumes a cheap, deterministic, and consistent comparator~\cite{graefe2006implementing}. LLM-based comparators violate these assumptions because generation is stochastic, outputs may depend on presentation order~\cite{wang2024faireval}, and APIs typically return permutations or pairwise preferences rather than calibrated scores~\cite{dai2023uncovering,qin2024large}. Accordingly, prior work has explored pointwise, pairwise, and listwise ranking paradigms. Dai et al.~\cite{dai2023uncovering} compare these approaches and show that no single paradigm dominates across tasks. 

Within database systems, Zhao et al.~\cite{zhao2025accesspaths} formalize the \texttt{LLM-ORDER BY} operator and study multiple execution strategies, including pairwise Quicksort (LOTUS~\cite{patel2024semantic}) and listwise Merge Sort schedules. These schedules reuse earlier answers only through their fixed structure (pivot partitions or sorted runs); none maintains an explicit global record of the preferences revealed across calls, so relations implied by transitivity across different batches cannot be used to skip later calls. In contrast, our work treats revealed preferences as shared state and exploits transitive inference to reduce
redundant LLM calls. \textsf{TBS} further designs its three-round grid
so that each round compares largely disjoint sets of pairs, and uses
greedy packing to cover the remaining unresolved pairs.

ListK~\cite{listk2026} studies semantic sorting and top-$K$ selection using fine-tuned listwise rankers such as RankZephyr~\cite{pradeep2023rankzephyr}. Whereas ListK focuses on specialized ranking models and pivot-based recursive schedules, our work targets general-purpose chat models and focuses on the execution of \texttt{LLM-ORDER BY} through preference-graph maintenance, transitivity-aware query skipping, and robustness to inconsistent outputs. 

SortBench~\cite{sortbench2025} evaluates the ability of LLMs to sort lists across numeric, lexical, and mixed ordering tasks. Our evaluation complements SortBench by studying semantic ordering as a database operator, measuring both ranking quality and execution cost across structured sorting, passage ranking, and recommendation workloads.

\subsection{Inconsistency of LLM Orderings}
\label{sec:related:consistency}

LLM-generated rankings are often inconsistent. Wang et al.\ show that GPT-4 exhibits position bias in pairwise comparisons, leading to unstable preference judgments~\cite{wang2024faireval}. Zeng et al.\ identify transitivity violations in LLM-generated rankings and propose LLM-RankFusion, which improves robustness by aggregating rankings from multiple rankers~\cite{zeng2024rankfusion}. These efforts aim to mitigate inconsistency through ranking aggregation and calibration. In contrast, our work treats inconsistency as a systems challenge for semantic sorting operators: detected contradictions are prevented from entering the maintained preference graph, ensuring that the stored ordering remains acyclic even when individual LLM outputs are noisy.

\subsection{LLM-Based Ranking and Reranking}
\label{sec:related:ranking}

LLMs are increasingly used as rerankers in retrieval pipelines, where an initial retrieval stage produces a candidate set that is subsequently reordered by an LLM~\cite{sun2023chatgpt,ma2023listwise,qin2024large}. These approaches primarily optimize top-$K$ relevance and retrieval metrics. Setwise prompting further reduces inference cost by ranking small subsets of candidates at a time~\cite{zhuang2024setwise}. In contrast, we study semantic \texttt{ORDER BY} execution, whose goal is to produce a complete ordering over a relation rather than rerank a fixed set of retrieved documents. Consequently, our focus is on reducing LLM comparison costs while maintaining a globally consistent ordering across the entire dataset.

\subsection{Learning to Rank}
\label{sec:related:learning2rank}

Learning-to-rank (LTR) methods train ranking functions using labeled relevance judgments or implicit feedback~\cite{liu2009learningtorank,joachims2002optimizing}. Classical pointwise, pairwise, and listwise approaches, including LambdaMART and modern cross-encoder architectures, remain highly effective when sufficient training data are available~\cite{burges2010from,nogueira2019passage}. However, LTR assumes a fixed ranking objective known during training. Semantic \texttt{ORDER BY} queries instead allow users to specify ranking criteria dynamically in natural language. Our goal is therefore not to learn a specialized ranker, but to execute such queries directly using a general-purpose LLM while minimizing comparison costs.

\subsection{Positioning of \textsf{TBS}}
\label{sec:related:positioning}

Prior work on LLM-based ranking spans pointwise, pairwise, and listwise
prompting, as well as database operators for semantic sorting, each trading
ranking quality against inference cost. As
Table~\ref{tab:related-compare} shows, systems that return a full order for
a supplied key list (OrderByLLM, LOTUS, ListK) do not maintain an explicit global record of the preferences revealed by earlier calls, while methods that address ranking inconsistency or quality (LLM-RankFusion, rerankers, Setwise) do not target
a database \texttt{ORDER BY}. Among the LLM-based methods in the table, \textsf{TBS} is the
only one that combines a global total order with a maintained preference
graph and transitivity-based call skipping.
Transitive reuse is established in crowdsourced sorting~\cite{marcus2011humanpowered}; our contribution is to make it work for listwise LLM calls at million-key scale. Additional related work is discussed in the Appendix.

\textsf{TBS} therefore addresses a complementary execution problem: reusing
preferences across listwise calls to avoid redundant LLM evaluations. It
records accepted preferences in an acyclic graph, uses reachability to
identify resolved comparisons, and stores only direct edges to keep memory
low. It also validates malformed replies and rejects cycle-inducing updates,
and we study how batch size and batch membership affect quality and cost.
Because the graph is built from LLM outputs, the benefit of reuse depends on
the accuracy of the accepted preferences, which makes the quality--cost
trade-off central to our evaluation.

\section{Preliminaries}
\label{sec:prelim}

\spara{LLM Sorting Operator.}
We model an LLM-based ordering operator, \texttt{LLM-order by}, that takes a list of $n$ keys $X$ and a natural-language ordering criterion $c$, and returns a ranked list consistent with $c$, namely a total order $\pi$ after deterministic tie-breaking. Keys may represent arbitrary objects, such as database row identifiers, document IDs, or named entities. The criterion is expressed in free text (e.g., ``sort by height ascending; break ties alphabetically'' or ``rank passages by relevance to the query''). In Example~\ref{ex:sitrep}, the keys are report \texttt{id} values and the criterion is the natural language urgency specification used in \texttt{LLM-order by}. The operator is accessed solely through a generative interface: given any subset $B \subseteq X$ and the criterion, the LLM returns an ordered list containing exactly the elements of $B$.

\spara{Extracting Pairwise Relations from Batch Outputs.}
For a batch size $m \ge 2$, let $B=\{b_1,\dots,b_m\}$ be a batch of keys. The LLM returns a permutation $\sigma(B) = (x_1,\dots,x_m)$ of $B$, which we interpret as the set of strict pairwise relations $x_i \succ x_j \quad \forall 1 \le i < j \le m$, i.e., the total order on $B$ implied by the listed sequence. This interpretation applies uniformly across tasks: in value-based sorting it captures numeric or ordinal comparisons, and in passage ranking it captures pairwise relevance preferences with respect to a query.

\spara{Maintained Relation Set, Closure, and Consistency.}
The algorithm maintains a set $\mathcal{R}$ of ordered pairs $(a,b)$, each indicating that $a$ is strictly preferred to $b$ under the criterion. After each batch response, all implied pairwise relations are added to $\mathcal{R}$, and the set is immediately closed under transitivity. Any relation already implied by $\mathcal{R}$ is skipped in future queries.

When inserting a new pair $(x,y)$, we propagate its transitive consequences using the reflexive sets $A=\{x\}\cup\{a : a \succ x\}$ and $S=\{y\}\cup\{b : y \succ b\}$ (equivalently, $A=\{x\}\cup\texttt{Greater\_than}(x)$ and $S=\{y\}\cup\texttt{Less\_than}(y)$ in \S\ref{sec:method:oracle}), thereby inferring every pair in $A\times S$. If the insertion would contradict existing relations, i.e., create a directed cycle, it is rejected so that $\mathcal{R}$ remains a strict partial order. Although the three-round schedule fixes batch membership by the grid layout, before each listwise call we test whether every pair in the scheduled batch is already resolved by closure, and skip the call if so. When a batch is queried because it still contains unresolved pairs, the model may return an ordering that contradicts relations already in $\mathcal{R}$; such edges are discarded before insertion.

\spara{Graph View and Incremental Indexing.}
We represent $\mathcal{R}$ as a directed graph $G=(V,E)$, where $V$ is the set of keys and an edge $u \!\to\! v$ indicates $(u,v)\in\mathcal{R}$. Since cycle-inducing updates are rejected, $G$ remains acyclic and therefore encodes a strict partial order. To support efficient updates, each key $x$ maintains two indexed sets:
\[
\begin{aligned}
  \texttt{Less\_than}(x) &= \{\, y : (x,y)\in\mathcal{R} \,\}, \\
  \texttt{Greater\_than}(x) &= \{\, y : (y,x)\in\mathcal{R} \,\}.
\end{aligned}
\]
When a new relation $(a,b)$ is accepted, we update these sets along the predecessors of $a$ and the successors of $b$, thereby incorporating all transitive consequences of the insertion.

\spara{Extracting a Total Order.}
Let $\mathrm{in}(x)=|\texttt{Greater\_than}(x)|$ denote the in-degree of $x$ in $G$. We obtain a linear order by performing a topological traversal, such as Kahn's algorithm~\cite{kahn1962topological}: repeatedly select a vertex with $\mathrm{in}(x)=0$, append it to the output, remove it for the purpose of degree updates, and decrement the in-degrees of its out-neighbors. When multiple vertices have in-degree zero, we apply a fixed deterministic tie-break rule (e.g., lexicographic order), which selects one linear extension of~$G$ (the lexicographically smallest topological order under a min-heap). The extension is unique only when~$G$ is already a total order. 

\spara{Problem Statement.}
We formalize the physical execution task addressed in this paper.
\begin{myProblem}[Transitivity-aware batched LLM sorting]
\label{prob:sort}
\textbf{Input:} A key list $X=[x_1,\ldots,x_n]$, a natural-language ordering criterion~$c$, a batch size $m\ge 2$, and a listwise sorter \textsf{LLM\_Sort} that, for any batch $B\subseteq X$ with $2\le |B|\le m$, returns a permutation $\sigma(B)$ of~$B$.

\textbf{Output:} A deterministic total order $\pi$ over~$X$.

\textbf{Interaction:} The algorithm selects batches $B_1,\ldots,B_T$ ($|B_t|\le m$), collects \textsf{LLM\_Sort} outputs $\sigma(B_t)$ by querying these batches, and inserts the resulting pairwise relations into a relation graph~$G$ under transitive closure. Batches whose pairs are already resolved are skipped, whereas any edge that would introduce a cycle in $G$ is excluded. When the schedule terminates, $\pi$ is extracted from~$G$ using Kahn’s topological sort with lexicographic tie-breaking among incomparable keys.

\textbf{Objective:} Minimize the number of listwise calls~$T$ while ensuring $\pi$ aligns closely with the task reference order (Kendall’s~$\tau$ for total-order tasks, nDCG for passage ranking; see \S\ref{sec:experiments}). Pairwise sorting (e.g., quicksort with an LLM comparator, QS$_1$) needs $\Theta(n\log n)$ calls, each settling one pair.
A listwise call on $m$ keys orders up to $\binom{m}{2}$ pairs at once, so our three-round batched schedule (\S\ref{sec:method}) needs $O(\tfrac{n}{m}\log_m n)$ calls, matching the lower bound for listwise sorting (\S\ref{sec:method:cost}), and transitivity-based skipping lowers the realized count further.
\end{myProblem}

\section{Methodology}
\label{sec:method}

This section presents {\em Transitive Batched Sort} (\textsf{TBS}) in three parts: the algorithm under a perfect listwise oracle (\S\ref{sec:method:oracle}); a direct-edge reachability index that avoids the quadratic memory of explicit transitive closure (\S\ref{sec:method:dag}); and a validation layer that converts raw LLM responses into permutations (\S\ref{sec:method:validation}). Heuristic optimizations for cost-efficient deployment of \textsf{TBS} in practice are presented in \S\ref{sec:optimization}.

\subsection{Transitive Batched Sort}
\label{sec:method:oracle}

\subsubsection{Oracle and Shared State}
A \emph{perfect oracle} returns, for any batch $B$ with $2 \le |B| \le m$,
the true permutation of $B$ under criterion $c$. \textsf{TBS} maintains three
structures that every procedure reads and updates:
\begin{enumerate}
  \item the relation graph $G = (V, E)$, kept closed under transitivity;
  \item the \emph{unresolved neighborhood} of each key, i.e., the set of keys
        whose relative order to $x$ is not yet known in $\mathcal{R}$
        (we write $x \mathbin{?} y$ when neither $x \succ y$ nor $y \succ x$ are known),
        \begin{equation}
        \label{eq:unknown-nbhd}
          U[x] \;=\; \{\, y \neq x : x \mathbin{?} y \,\}. \nonumber
        \end{equation}
  \item the number of unresolved unordered pairs,
        \begin{equation}
        \label{eq:unknown-count}
          u \;=\; \tfrac{1}{2} \sum_{x \in X} |U[x]|. \nonumber
        \end{equation}
\end{enumerate}
Initially, $E = \emptyset$, $U[x] = X \setminus \{x\}$ for all $x \in X$, and
$u = \binom{n}{2}$.

Logically, $G$ is a total order if and only if $u = 0$. The closed graph $G$, the sets $U[\cdot]$, and the counter $u$ serve only as a \emph{logical} specification of the algorithm, since materializing them explicitly requires $\Theta(n^2)$ storage, which is prohibitive at scale. In practice they are never materialized; they are derived on demand from a sparse direct-edge DAG (\S\ref{sec:method:dag}), which keeps memory linear in $n$ and the number of stored preferences. As a consequence, the test $u = 0$, which is $O(1)$ under explicit closure with a maintained counter, is no longer $O(1)$ in the implementation, where $u$ is recovered by blocked column sweeps.

\subsubsection{Batch Grid}
\label{sec:method:overview}
For a key list $X$ and $k=\lceil |X|/m\rceil$, \textsf{TBS} splits $X$ into columns $C_0,\ldots,C_{k-1}$ of at most $m$ consecutive keys.
Let $C_j[r]$ be the $r$-th key of $C_j$. Row $r$ is $R_r=\{C_j[r] : 0\le j<k,\ r<|C_j|\}$. Diagonal $d$ ($d=0,\ldots,m{+}k{-}2$, but may be fewer if the last column is
incomplete) is
$D_d = \bigl\{\, C_j[r] \;\big|\; r + j = d,\; 0 \le j < k,\; r < |C_j| \,\bigr\}$,
so $|D_d|\le\min(m,k)$.
A diagonal with fewer than two keys contains no
pair to compare and is never queried. This includes $D_0=\{C_0[0]\}$ and, when
the last column is full, $D_{m+k-2}=\{C_{k-1}[m-1]\}$.

A useful way to read the grid is by pair families: columns cover intra-column comparisons, rows propagate order across columns at matched ranks, and diagonals cover the residual cross-rank boundary interactions left after the first two rounds.

\subsubsection{Algorithm}
\textsf{TBS} is organized as three algorithms that operate on a shared state
(the relation graph $G$, the unresolved neighborhoods $U[\cdot]$, and the
counter $u$). Algorithm~\ref{alg:oracle_tbs} is the main procedure: under a
perfect oracle it runs the three-round schedule of
Algorithm~\ref{alg:schedule} once on~$X$, then calls \textsc{Pack} once to
order the remaining unresolved pairs, after which $u=0$ and the topological
order of $G$ is unique. Algorithm~\ref{alg:helpers} defines the two primitives
through which oracle calls are issued: \textsc{Resolve}, which sorts one batch
with a single oracle call (skipped when every pair in the batch is already
ordered) and merges the answer into $G$, and \textsc{Pack}, which builds
batches for the pairs left unresolved and sends them to \textsc{Resolve}.

\begin{algorithm}[t]
\footnotesize
\caption{\small{Perfect Oracle Transitive Batched Sort}}
\label{alg:oracle_tbs}
\begin{flushleft}
\textbf{Input:} key set $X$ with $n=|X|$, ranking criterion $c$, batch size $m\ge 2$\\
\textbf{Output:} total order $\pi$ over $X$\\
\textbf{Shared state:} relation graph $G=(V,E)$ over $V=X$, unresolved neighborhoods $U[\cdot]$, and unresolved-pair counter $u$, read and updated by Algorithms~\ref{alg:schedule} and~\ref{alg:helpers}
\end{flushleft}
\footnotesize
\begin{algorithmic}[1]
\STATE $G \gets (V=X, E=\emptyset)$
\STATE $U[x] \gets X\setminus\{x\}$ for all $x\in X$
\STATE $u \gets \binom{n}{2}$
\STATE \textsf{Schedule}(X) \COMMENT{three-round recursive schedule}
\STATE \textsf{Pack}() \COMMENT{resolve remaining unresolved pairs}
\STATE \textbf{return} $\pi \gets \textsf{Topological\_Sort}(G)$
\end{algorithmic}
\end{algorithm}

\begin{algorithm}[t]
\footnotesize
\caption{\small{$\textsf{Schedule}(L)$: Three-round Schedule on Key List $L$}}
\label{alg:schedule}
\begin{flushleft}
\textbf{Input:} key list $L$\\
\textbf{Uses:} batch size $m$ and shared state $G$, $U[\cdot]$, $u$\\
\textbf{Effect:} updates $G$, $U[\cdot]$, and $u$
\end{flushleft}
\footnotesize
\begin{algorithmic}[1]
\IF{$u=0$ \textbf{or} $|L|<2$}
  \RETURN
\ENDIF
\IF{$|L|\le m$}
  \STATE \textsf{Resolve}($L$)
  \RETURN
\ENDIF
\STATE $k \gets \lceil |L|/m \rceil$
\STATE split $L$ into columns $C_0,\dots,C_{k-1}$, each of size at most $m$
\STATE \textbf{Round 1: Column sorting}
\FOR{$j=0$ \TO $k-1$}
  \STATE \textsf{Resolve}($C_j$); keep $C_j$ in sorted order
\ENDFOR
\STATE \textbf{Round 2: Row sorting with recursion}
\FOR{$r=0$ \TO $m-1$}
  \STATE $R_r \gets \{C_j[r] : 0\le j<k,\ r<|C_j|\}$
  \STATE \textsf{Schedule}($R_r$) \COMMENT{$|R_r|\le \lceil |L|/m\rceil < |L|$ for $|L|>m$}
\ENDFOR
\STATE \textbf{Round 3: Diagonal stitching}
\FOR{$d=1$ \TO $m+k-3$}
  \IF{$u=0$}
    \RETURN
  \ENDIF
  \STATE $D_d \gets \{C_j[r] : 0\le j<k,\ 0\le r<|C_j|,\ r+j=d\}$
  \IF{$|D_d|\ge 2$}
    \STATE \textsf{Resolve}($D_d$)
  \ENDIF
\ENDFOR
\end{algorithmic}
\end{algorithm}

\begin{algorithm}[t]
\footnotesize
\caption{\small Oracle-call Primitives: \textsf{Resolve} and \textsf{Pack}}
\label{alg:helpers}
\footnotesize
\begin{algorithmic}[1]
\STATE \textbf{procedure} \textsf{Resolve}($B$)
\IF{$|B| < 2$}
    \RETURN
\ENDIF
\IF{there exist $a,b\in B$ such that $b\in U[a]$}
    \STATE $\sigma \gets \mathsf{Oracle\_Sort}(B,c)$ \COMMENT{true order of batch $B$ under criterion $c$}
    \STATE insert into $G$ all pairwise relations implied by $\sigma$
    \STATE update $G$ incrementally under transitivity
    \STATE for each newly resolved unordered pair $\{x,y\}$, remove $y$ from $U[x]$ and $x$ from $U[y]$
    \STATE update $u$
\ENDIF
\end{algorithmic}

\vspace{2pt}
\hrule
\vspace{2pt}

\begin{algorithmic}[1]
\STATE \textbf{procedure} \textsf{Pack}()
\WHILE{$u>0$}
    \STATE $b \gets \arg\max_{x\in X}|U[x]|$
    \STATE $B \gets \{b\}$ \COMMENT{seed with the key of highest unresolved degree}
    \WHILE{$|B|<m$}
        \STATE $s \gets \max_{c\notin B}|U[c]\cap B|$
        \IF{$s=0$}
            \STATE \textbf{break}
        \ENDIF
        \STATE $d \gets \arg\max_{c\notin B}|U[c]\cap B|$
        \STATE $B \gets B\cup\{d\}$
    \ENDWHILE
    \STATE \textsf{Resolve}($B$)
\ENDWHILE
\end{algorithmic}
\end{algorithm}

Given a key list $X$, \textsf{Schedule} (Algorithm~\ref{alg:schedule}) issues a single \textsf{Resolve} call if $|X|\le m$. Otherwise, it splits $X$ into $k=\lceil |X|/m\rceil$ columns and proceeds in three rounds.
\emph{Round~1} resolves each column and keeps it in sorted order, yielding $k$ disjoint chains.
\emph{Round~2} forms each row $R_r$ from the $r$-th keys of the sorted columns and schedules it recursively on the same $G$, $U[\cdot]$, and $u$. Since $|R_r|\le k<|X|$, the recursion terminates. Because each sorted column is a chain with position~$0$ highest, a row answer $x\succ y$ with $x=C_i[r]$ and $y=C_j[r]$ implies, by transitivity, that every key at or above~$x$ in~$C_i$ precedes every key at or below~$y$ in~$C_j$. This yields $(r+1)(|C_j|-r)$ cross-column pairs, which, over all~$r$, is at most $\bigl(\tfrac{m+1}{2}\bigr)^2$ when $|C_j|=m$.
\emph{Round~3} resolves the diagonals $D_1,\ldots,D_{m+k-3}$. Keys on a diagonal lie in pairwise different rows and columns, so Rounds~1--2 never compare them directly; each diagonal call stitches adjacent rows across column boundaries. The schedule returns as soon as $u=0$.

\textsf{Resolve}$(B)$ first checks $U[\cdot]$ for an unresolved pair inside the
batch~$B$; if there is none, no call is issued, so batches already settled by
transitivity cost nothing. Otherwise, one oracle call returns
$\sigma(B)=(x_1,\ldots,x_{|B|})$, whose edges $x_i\succ x_j$ for $i<j$ are
merged one at a time. An edge $a\succ b$ is \emph{skipped} if
$b\in\texttt{Less\_than}(a)$, since it is already known, and \emph{rejected} if
$a\in\texttt{Less\_than}(b)$, since it would close a cycle; rejection cannot
occur under the oracle, whereas under an LLM the edge is discarded and counted
as a contradiction. Otherwise, let
$\mathcal{A}=\{a\}\cup\texttt{Greater\_than}(a)$ and
$\mathcal{S}=\{b\}\cup\texttt{Less\_than}(b)$. By transitivity, every pair
$(a',b')\in\mathcal{A}\times\mathcal{S}$ not yet in~$E$ is inserted, $a'$ and
$b'$ are removed from each other's neighborhoods, and $u$ is decremented once
per pair. Since each unordered pair leaves $U[\cdot]$ at most once, the amortized cost of
\emph{removals} is $O(n^2)$ over the whole run.
Locating which pairs of $\mathcal{A}\times\mathcal{S}$ are new, however, costs
$O(|\mathcal{A}|\,|\mathcal{S}|)$ per merge (up to $O(n^2)$ in the worst case per
call), so without a structure that enumerates only new pairs the total scan cost
can reach $O(T\,n^2)$; \S\ref{sec:method:dag} gives the efficient
direct-edge implementation used in practice.

The diagonal phase need not resolve every cross-pair. For example, the relations $a \succ b$, $a \succ c$, $b \succ d$, and $c \succ d$ leave the order of $b$ and $c$ unknown if no batch contains both keys. Such residual comparisons are exactly those recorded in the unresolved neighborhoods $U[\cdot]$. To resolve them, \textsf{Pack} greedily builds each batch: it seeds the batch with the key of highest unresolved degree, then repeatedly adds the key that shares the most unresolved pairs with the current batch. Every packed batch therefore contains at least one unresolved pair, so each call to \textsf{Resolve} strictly decreases $u$ under a perfect oracle, which guarantees termination. Because \textsf{Pack} is driven by $U[\cdot]$, it examines only the leftover unresolved pairs and never rescans the full $\binom{n}{2}$ pair space.

\subsubsection{Correctness}
Under the oracle, every accepted edge is correct and no insertion is rejected, so $G$ remains a transitively closed subgraph of the true order. Since \textsf{Pack} terminates only when $u=0$, every pair of keys is eventually ordered, and a topological sort of $G$ therefore recovers the true total order ($\tau = 1$). Under a noisy comparator, cycle-inducing edges are rejected, preserving the acyclicity of $G$. Consequently, the output is always a valid linear extension of the accepted preferences, with remaining incomparable keys ordered by a deterministic lexicographic tie-break rule.

\subsubsection{Cost}
\label{sec:method:analysis}
\label{sec:method:cost}
We measure operator cost as the number of listwise LLM calls, counting each batch request as a single call regardless of its size ($2 \le |B| \le m$).
Because calls differ in length, we also report total tokens and USD cost for every run (\S\ref{sec:exp:metrics}).
Ignoring transitivity-based skipping, the call complexity of \textsc{Schedule} (Algorithm~\ref{alg:schedule}, all rows) on an input of size $n$ satisfies
\[
T(n) \le m\,T\!\left(\left\lceil \frac{n}{m} \right\rceil\right) + O\!\left(\frac{n}{m} + m\right),
\]
where Round~1 issues at most $\left\lceil n/m \right\rceil$ column calls, Round~2 recursively sorts $m$ rows of size at most $\left\lceil n/m \right\rceil$, and Round~3 evaluates at most $\left\lceil n/m \right\rceil + m - 1$ diagonal batches.
Level $i$ of the recursion has $m^{i}$ subproblems of size $n/m^{i}$, each issuing $O(n/m^{i+1}+m)$ calls, so level $i$ issues $O(n/m + m^{i+1})$ calls.
Summing over the $\log_m n$ levels, the first term gives $O\!\left(\frac{n}{m}\log_m n\right)$ and the second, a geometric series dominated by the deepest level, gives $O(n)$. Hence
\[
T(n) = O\!\left(\frac{n}{m}\log_m n + n\right).
\]
Each call on $m$ keys reveals at most $\log_2 m!$ of the $\log_2 n!$ bits needed, so any listwise scheme requires $\Omega(\log n! / \log m!) = \Omega((n/m)\log_m n)$ calls; the leading term of \textsc{Schedule} matches this bound.
The bound concerns \textsc{Schedule} only; \textsc{Pack} that guarantees completeness, adds a few more %at most $u'$ 
calls.

%The $R_0$-only default (\S\ref{sec:method:optimizations}) recurses on a single row, $T(n) \le T(\lceil n/m \rceil) + O(n/m + m)$, so its schedule issues $O(n/m + m\log_m n)$ calls. For example, $2\lceil n/m \rceil + m = 50$ for $n{=}200$ and $m{=}10$, in line with the roughly $50$ non-\textsc{Pack} calls observed on NBA and population.

%The $R_0$-only schedule leaves more pairs unresolved, and \textsc{Pack} orders them. After the schedule, let $u'$ be the number of unresolved pairs. Each packed batch contains at least one such pair, so \textsc{Pack} adds at most $u'$ calls, and typically far fewer because each call also resolves pairs implied by transitivity. In our experiments \textsc{Pack} accounts for $33$--$50\%$ of all calls (Table~\ref{tab:rq2-inconsistency}), so the total is $1.5$--$2\times$ the schedule's call count; it is included in every call count we report. Transitivity-based skipping and early stopping reduce the realized count below these bounds.

\subsection{Memory-Efficient Reachability Index}
\label{sec:method:dag}
 
Explicit transitive closure (\S\ref{sec:method:oracle}) stores every known and every open pair, so its memory is $\Theta(n^2)$ regardless of progress, and merging an edge $a\succ b$ inserts one pair for every ancestor of $a$ and descendant of $b$ (including $a$ and $b$ themselves), up to $n^2/4$ pairs per merge\footnote{\scriptsize Let $\mathcal{A}$ be $a$ together with its ancestors and $\mathcal{S}$ be $b$ together with its descendants, so a merge inserts $|\mathcal{A}|\cdot|\mathcal{S}|$ pairs. The two sets are disjoint: a key in both would give $b\leadsto a$, which is rejected before insertion. Hence $|\mathcal{A}|+|\mathcal{S}|\le n$, and by the AM--GM inequality \cite{hardy1952inequalities}, $|\mathcal{A}|\cdot|\mathcal{S}|\le\big(\tfrac{|\mathcal{A}|+|\mathcal{S}|}{2}\big)^2\le n^2/4$.}.
We instead store only the accepted answers as a sparse DAG and derive all closure quantities on demand, so memory scales with the number of answers rather than with $n^2$, and each insertion touches only the affected part of the graph.
The index is purely an implementation change: it issues the same batches, accepts and rejects the same edges, and produces the same output $\pi$ as explicit closure.
 
\subsubsection{Direct Edges}
Let $H=(X,E_d)$ contain only the edges accepted from listwise answers, and write $x\leadsto y$ if $H$ has a directed path of length at least one\footnote{\scriptsize A zero-length path would give $x\leadsto x$ and hence $x\succ_G x$, i.e., a key ranked above itself.} from $x$ to $y$.
The closure $G$ is then defined by $x \succ_G y \Longleftrightarrow x \leadsto y$.

\textsf{Resolve}$(B)$ examines the pairs of $\sigma(B)=(x_1,\ldots,x_{|B|})$ in a fixed order: consecutive pairs $(x_i,x_{i+1})$ first, then $(x_i,x_j)$ with $j\ge i{+}2$.
A pair $(a,b)$ is \emph{skipped} if $a\succ_G b$, \emph{rejected} as a contradiction if $b\succ_G a$, and otherwise appended to $E_d$.
If no consecutive pair is rejected, the path $x_1\to\cdots\to x_{|B|}$ lies in $G$, so all remaining pairs are already implied and are skipped without testing. Otherwise, the descendants and ancestors of each $x_i$ are marked once by a pruned search, and the marks are updated after each insertion, which yields the same outcome as testing every pair individually. Because acceptance depends only on reachability, which is exactly the relation stored by explicit closure, both implementations make identical decisions. The memory is $O(n+|E_d|)$, where $E_d$ holds at most one edge per consecutive pair of each answer, plus a few non-consecutive insertions.
 
\subsubsection{Dynamic Topological Order}
To answer reachability queries without storing the closure, we maintain a topological order of $H$ that bounds the search region of each query and is repaired locally after each insertion. In particular, we maintain a bijection $\mathrm{ord}:X\to\{0,\ldots,n-1\}$ such that
$(a,b)\in E_d \;\Longrightarrow\; \mathrm{ord}(a)<\mathrm{ord}(b)$, and repair it after each insertion with the algorithm of Pearce and Kelly~\cite{pearce2006dynamic}.
Inserting $(a,b)$ with $\mathrm{ord}(a)<\mathrm{ord}(b)$ requires no change.
Otherwise, only the affected region is reordered, where $\leadsto$ is taken to be reflexive for this definition:
\begin{equation}
\label{eq:dag-pk}
  \begin{aligned}
  \delta_F&=\{v : b\leadsto v,\ \mathrm{ord}(v)<\mathrm{ord}(a)\},\\
  \delta_B&=\{v : v\leadsto a,\ \mathrm{ord}(v)>\mathrm{ord}(b)\}.
  \end{aligned}
  \nonumber
\end{equation}
The positions $\mathrm{ord}(\delta_B\cup\delta_F)$ are pooled and reassigned, first to $\delta_B$ and then to $\delta_F$, each in its previous relative order.
The two sets are disjoint because $b\not\succ_G a$ was verified before the insertion.
The cost is $O\big(\sum_{v\in\delta_B\cup\delta_F}(1+\deg v)+|\delta|\log|\delta|\big)$, with $\delta=\delta_B\cup\delta_F$, which is local to the region.
Any valid initial order yields the same answers, so $\mathrm{ord}$ may be seeded from a heuristic ranking to reduce reordering.
 
\subsubsection{Pruned Reachability Queries}
Every vertex on a path from $x$ to $y$ precedes $y$ in $\mathrm{ord}$ (or equals $y$), hence
\begin{equation}
\label{eq:dag-query}
  x\succ_G y \;\Longleftrightarrow\; \mathrm{ord}(x)<\mathrm{ord}(y)\ \wedge\ y\in\mathrm{Reach}_H(x;\,S_y), \nonumber
\end{equation}
where $S_y=\{v:\mathrm{ord}(v)<\mathrm{ord}(y)\}\cup\{y\}$ and $\mathrm{Reach}_H(x;S)$ is the set of vertices reached by a depth-first search from $x$ that expands only vertices in $S$. If $\mathrm{ord}(x)>\mathrm{ord}(y)$, the query returns false without any search.

\spara{Chain Shortcuts.} Round~1 sorts each column $C_j$ with one listwise call, so every column is a
chain in $G$: it holds the densest block of known relations ($\binom{|C_j|}{2}$
pairs) but requires only $|C_j|{-}1$ stored edges. Columns are disjoint and
sorted before any other call, so their answers cannot contradict $G$ and are
accepted in full. We exploit this structure with a fixed chain cover that
stores, for each key $x$, its column $\mathrm{ch}(x)$ and its position
$\mathrm{pos}(x)$ within it (position~$0$ is the highest-ranked key), $2n$
integers in total. Labeling vertices by chain and position and answering
queries by per-chain comparisons is a classical technique for compressing
transitive closure~\cite{Jagadish90,Simon88}. In our setting the cover comes at
no construction cost: the Round-1 columns are already sorted chains, so no
chain decomposition is computed, and the cover stays fixed as edges are
inserted.

These labels speed up our queries in two ways:
\begin{enumerate}
  \item \emph{Same chain.} If $\mathrm{ch}(x)=\mathrm{ch}(y)$, the column order settles the query without any search: $x\succ_G y\Leftrightarrow\mathrm{pos}(x)<\mathrm{pos}(y)$.
  \item \emph{Early termination.} If the search from $x$ reaches a vertex $v$ with $\mathrm{ch}(v)=\mathrm{ch}(y)$ and $\mathrm{pos}(v)\le\mathrm{pos}(y)$, then $v=y$ or $v\succ_G y$ by the chain order. Together with $x\leadsto v$, transitivity gives $x\succ_G y$, so the search stops with a positive answer without locating $y$ itself.
\end{enumerate}
Both rules shorten the search and never change its outcome. 

\spara{Batch Skip Test.}
Before issuing an oracle call, \textsf{Resolve}$(B)$ tests whether $B$ still contains an unresolved pair (Algorithm~\ref{alg:helpers}, line~4).
Explicit closure answers this by scanning $U[\cdot]$. Since $U[\cdot]$ is not materialized here, we instead check whether every pair $\{a,b\}\subseteq B$ is already ordered in $G$.
Because $x\succ_G y$ requires $\mathrm{ord}(x)<\mathrm{ord}(y)$, each pair needs a query in only one direction, the one consistent with $\mathrm{ord}$, and the chain shortcuts apply.
The check stops at the first pair that is ordered in neither direction, so it usually terminates early. If every pair is ordered, the batch is skipped at no cost.

\spara{Exact Counts by Blocked Column Sweeps.} 
The scheduler repeatedly requires two statistics. The unresolved degree $d(x) = |U[x]|$ is used by \textsf{Pack} to seed each batch with the key of highest unresolved degree (Algorithm~\ref{alg:helpers}, \textsf{Pack}, line 3). The total number of unresolved pairs $u$ is used in the termination tests of \textsf{Schedule} (Algorithm \ref{alg:schedule}, lines 1 and 17) and in the loop condition of Pack (Algorithm~\ref{alg:helpers}, \textsf{Pack}, line 2). Under the explicit-closure formulation, these quantities are available directly from the maintained ancestor and descendant sets. In the direct-edge implementation, however, the transitive closure is never materialized. We therefore recover the same counts directly from the DAG $H=(X,E_d)$ using a blocked column sweep. Recall that the Round-1 columns $C_0,\ldots,C_{w-1}$ form a chain cover of $H$. For a block of columns, we perform a reverse sweep in topological order. For each key $x$ and column $c$, we maintain a transient label \[
\begin{aligned}
\lambda[x][c] = \min\Big(
  &\{pos(x)+1 : ch(x)=c\} \\
  &\cup \{\lambda[y][c] : (x,y)\in E_d\} \\
  &\cup \{pos(y) : (x,y)\in E_d,\ \mathit{ch}(y)=c\}
\Big),
\end{aligned}
\] with $\min\emptyset = +\infty$.
The middle set propagates $\lambda[y][c]$ from \emph{every} successor~$y$, whatever column~$y$ lies in; only the direct-position term is guarded by $\mathit{ch}(y)=c$. The label identifies the first position in column $c$ that is known to be reachable from $x$: \[ \lambda[x][c] = \min\{p : x \succ_G C_c[p]\}. \] Since descendants of $x$ within a chain form a suffix of that chain, the number of descendants of $x$ is \[ |D(x)| = \sum_{c=0}^{w-1} \max\!\bigl(0,\,|C_c|-\lambda[x][c]\bigr). \] A symmetric forward sweep computes $|A(x)|$. The unresolved degree then follows directly as \[ d(x) = (n-1)-|D(x)|-|A(x)|. \] Summing $|D(x)|$ over all keys yields the total number of ordered pairs; the number of unresolved pairs is the remaining complement. Thus, the blocked sweep computes the exact statistics required by the scheduler without materializing the transitive closure.
 
\spara{Completeness Test.}
Deciding whether $G$ is a total order requires no count. Since $\mathrm{ord}$ is a
topological order of $H$, $G$ is total iff every pair of consecutive keys
is a direct edge:
\[
\footnotesize
\textsc{Complete}(G)\Leftrightarrow
\big(\mathrm{ord}^{-1}(i),\mathrm{ord}^{-1}(i{+}1)\big)\in E_d\ \ \forall\, i<n{-}1.
\]
Indeed, if $G$ is total, consecutive keys form covering pairs, which must be direct
edges; conversely, the listed edges form a Hamiltonian path, which implies all pairs.
The test costs $O(n+|E_d|)$ in the worst case and typically fails at the first
missing edge.
 
\subsubsection{Packing and Output}
\textsf{Pack} seeds each batch with $\arg\max_x d(x)$, obtained from one pair of
sweeps. To extend a batch $B$, it needs the coverage gain $|U[c]\cap B|$ of every candidate $c$, which is maintained incrementally: after
adding $b'$ to $B$, a forward and a backward search from $b'$ mark its descendants
and ancestors $D(b')\cup A(b')$, i.e., the keys already ordered with $b'$, and
every unmarked candidate gains one. This costs $O\big(m\,(n+|E_d|)\big)$ per batch.
Ties are broken by lexicographic key order.
The final order is obtained by running Kahn's algorithm on $H$ with a min-heap
keyed by lexicographic rank, in $O\big((n+|E_d|)\log n\big)$ time. A key becomes
ready exactly when all its ancestors in $G$ have been emitted, so the output equals
the topological order of $G$ with lexicographic tie-breaking.

\subsection{LLM Output Validation and Repair}
\label{sec:method:validation}

A listwise reply is not guaranteed to be a permutation of the requested
batch. Before merging, TBS parses the reply into a list~$L$, normalizes
item identifiers by case-insensitive matching against the batch~$B$, and
verifies that $L$ is a permutation of $B$ (Table~\ref{tab:validation-checks}).
If any check fails, $L$ is repaired by keeping only keys that appear in both
the reply and~$B$, in reply order (first occurrence of each), and dropping
tokens that do not match~$B$.
A reply that omits keys is re-asked, up to five attempts.
If it is still incomplete, the missing keys are appended in presentation order; in our runs this fallback never fired.
If fewer than two valid keys remain, the reply is discarded and no edges are
merged from that call.
Repair therefore preserves the model's relative order on the keys it actually
ranked, and contributes only those edges.

Validation addresses \emph{format} errors only. A well-formed reply that
contradicts the preference graph~$G$ passes validation and is instead
handled by cycle-inducing edge rejection at merge time.

\begin{table}[t]
  \centering
  \caption{\small Format validation checks for LLM listwise replies.}
  \label{tab:validation-checks}
  \scriptsize
  \setlength{\tabcolsep}{4pt}
  \renewcommand{\arraystretch}{1.1}
  \begin{tabular}{@{}l l p{0.36\columnwidth}@{}}
    \toprule
    \textbf{Check} & \textbf{Requirement} & \textbf{Typical violation} \\
    \midrule
    Parsing        & $|L| \ge 1$                & empty or unparseable reply \\
    Minimum length & $|L| \ge 2$ (if $|B| \ge 2$) & single-item list \\
    Completeness   & $B \subseteq L$            & omitted batch key \\
    No extras      & $L \subseteq B$            & hallucinated item \\
    No duplicates  & each key appears once      & repeated key \\
    Item matching  & case-insensitive match to $B$ & case or format mismatch \\
    \bottomrule
  \end{tabular}
\end{table}

\section{Optimization Strategies}
\label{sec:optimization}

Algorithm~\ref{alg:oracle_tbs} assumes a perfect oracle and runs until every
pair is ordered ($u=0$). With an LLM, each call is costly and replies may be
noisy, so fully resolving all $\binom{n}{2}$ pairs is not always worthwhile.
This section describes heuristics layered on Algorithm~\ref{alg:oracle_tbs}
that (i)~reduce the number of calls, (ii)~control which keys share a batch, and
(iii)~select the batch size~$m$. They change which batches are issued and when
the schedule stops, not how replies are merged into~$G$. Hence $G$ stays
acyclic and $\pi$ remains a linear extension of~$G$
(\S\ref{sec:method:oracle}). They are not part of the oracle correctness
argument, and they may leave $u>0$.

\subsection{Cost-Reducing Execution Rules}
\label{sec:method:optimizations}

\spara{Round-2 Row Sorting Options.}
Algorithm~\ref{alg:schedule} sorts every row recursively. To save calls, Round~2
may instead sort only the head row~$R_0$, either recursively or with
overlapping windows of at most $m$ keys. Deeper rows are then related to the
rest indirectly through diagonal stitching and transitivity.
We use recursive $R_0$-only as the default in our experiments; sorting every row (Algorithm~\ref{alg:schedule}) is evaluated as an ablation (Table~\ref{tab:round2-rows}).

\spara{Diagonal Batching.}
Short diagonals have few keys. A call on one of them uses a small
part of the batch capacity. Hence, consecutive short diagonals are merged
into a single batch of at most $m$ keys, while longer diagonals are issued
alone.

\spara{Early Stopping.}
\label{sec:method:stopping}
Let $u_t$ be the number of unresolved pairs after the $t$-th Round-3 batch and
$\rho_t = 1 - u_t/\binom{n}{2}$ the fraction of ordered pairs.
Round~3 (diagonals, then Pack) stops issuing calls as soon as one of three conditions holds: \emph{completion}
($u_t=0$); \emph{stall}, when a fixed number of consecutive executed batches
add no new ordered pair; or \emph{$\theta$-completion}, $\rho_t \ge \theta$ for
a target $\theta\in(0,1]$.
If $\rho_t \ge \theta$ is reached during the diagonals, the remaining diagonals are skipped; otherwise Pack continues until it is reached.
Stopping inserts no edge, so $\pi$ remains a linear
extension of~$G$. The $u_t$ pairs still unresolved are ordered by the
deterministic tie-break of \S\ref{sec:prelim}, so stopping affects only the
order of these pairs.

\spara{Unresolved Packing.}
Under the oracle, Algorithm~\ref{alg:oracle_tbs} always ends with
\textsf{Pack}, which is required for $\tau=1$. Under an LLM, \textsf{Pack} is
an optional switch. It spends additional calls on the residual pairs, and
ordering them with noisy replies can add contradictions. We enable it when
completeness is preferred over cost and otherwise rely on early stopping.

% \spara{Embedding Cascade.}
% For relevance ranking, a cheap embedding-based scorer can resolve some pairs
% without an LLM call. Such pairs are added to~$G$ through the same merge
% routine as LLM preferences, so a batch whose pairs are all resolved this way is
% skipped.

\subsection{Batch Membership and Prompt Order}
\label{sec:method:batchmembership}

Listwise quality depends not only on $m$ but also on which keys share a batch
and in what order they appear in the
prompt~\cite{wang2024faireval,tang2024permutation}. In TBS, membership is fixed
by the Round-1 partition of $X$ into columns $\{C_j\}$. The rows (Round~2) and
diagonals (Round~3) are derived from the same partition, so a single
\emph{membership preset} controls the entire batch grid. We consider four
presets:
\begin{itemize}[leftmargin=*]
  \item \textbf{Contiguous}: split $X$ into consecutive segments of size~$m$
        in input order (inputs are shuffled once before sorting, so this is already a random partition);
  \item \textbf{Random}: re-shuffle $X$ with a second seed before splitting; comparing it with contiguous measures how much results vary between two random partitions;
  \item \textbf{Similarity}: cluster keys by token-based textual similarity and
        pack clusters into columns of size~$m$;
  \item \textbf{Similarity\_embed}: same packing, but clusters use embedding
        similarity.
\end{itemize}
Keys are listed in partition order in every prompt. Which preset works best
depends on the task and model; we study this in our experiments.

\subsection{Batch-Size Selection}
\label{sec:method:batchsize}

The batch size $m$ trades the number of LLM API calls against quality. A small $m$ yields more
columns, rows, and diagonals, hence more calls. A large $m$ strains the context
window and degrades listwise accuracy. We therefore select $m$ per model and
task from a candidate set $\mathcal{M}$ with a one-time sweep. For each
$m\in\mathcal{M}$ we run TBS once and record
the total number of calls $c_m$ and a quality score $q_m$ (Kendall's~$\tau$ or
nDCG). The points $(c_m,q_m)$ are sorted by call count and re-indexed as
$(x_i,y_i)$, $i=0,\ldots,N{-}1$, with $N=|\mathcal{M}|$. The chosen $m$ is then
fixed for subsequent runs on that model and task. The sweep needs a reference
order to compute $q_m$; label-free estimation is left to future work.

\spara{Geometric knee.}
For $N\ge 3$, let $P_0$ and $P_{N-1}$ be the call-minimum and call-maximum
points. The \emph{knee} $i^\star$ is the interior point with the largest
perpendicular distance to the chord $\overline{P_0P_{N-1}}$~\cite{satopaa2011kneedle}.

\spara{Endpoint-aware utility.}
We normalize calls and quality to $[0,1]$,
\begin{align}
  \tilde{x}_i &= \frac{x_i - x_{\min}}{x_{\max}-x_{\min}}, &
  \tilde{y}_i &= \frac{y_i - y_{\min}}{y_{\max}-y_{\min}}, \nonumber
  \\
  U_i &= \alpha\,\tilde{y}_i + (1-\alpha)\,(1-\tilde{x}_i), &
  \alpha &\in [0,1], \nonumber
\end{align}
where a smaller $\alpha$ favors fewer calls.
We use $\alpha = 0.7$, weighting quality more than call savings.
If an endpoint ($i=0$ or $i=N{-}1$)
has strictly higher utility than the knee, we select that endpoint's batch
size. Otherwise we select $m_{i^\star}$.

\section{Experimental Results}
\label{sec:experiments}

We evaluate whether \textbf{Transitive Batched Sort (\textsf{TBS})} improves the quality--cost trade-off of LLM-based ordering by exploiting transitivity to reuse preferences across batched queries. We compare TBS against pairwise, pointwise, and listwise baselines; quantify the API-call savings from transitivity-aware optimizations; and analyze how key design choices affect ranking quality and cost. To assess robustness and generality, we evaluate \textsf{TBS} across multiple LLM providers and two benchmark families: total-order recovery (\emph{NBA heights}, \emph{World population}) and relevance-based ranking (\emph{DL19}, \emph{SciFact}, \emph{MovieLens}). \textbf{Our code and data are available at \url{https://github.com/kakinola23/TBS-LLM-OrderBy}}.

\spara{Research Questions.}
\begin{enumerate}
    \item[\textbf{RQ1}] \textbf{Quality--cost trade-off.} Can \textsf{TBS} match or exceed the ordering quality (Kendall~$\tau$, $\mathrm{nDCG}$) of the baseline methods  while using substantially fewer LLM API calls?
    
    \item[\textbf{RQ2}] \textbf{Contribution of optimizations.} How much does each design component (transitivity maintenance with query skipping, the reachability index, Round~2 row-sorting heuristics, and Round~3 early termination) contribute to reducing API calls and preserving ranking quality?
    
    \item[\textbf{RQ3}] \textbf{Robustness across models and tasks.} How do quality and cost vary across LLM providers and across the total-order and relevance-based benchmarks?
    
    \item[\textbf{RQ4}] \textbf{Batch-size sensitivity.} How does the batch size~$m$, together with the proposed knee/endpoint selection strategy, affect quality and cost?
    
    \item[\textbf{RQ5}] \textbf{Batch strategies.} How do Round~1 membership presets (\textsf{contiguous}, \textsf{random}, \textsf{similarity}, \textsf{similarity\_embed}) influence ranking quality and API-call count?
\end{enumerate}

\subsection{Experimental Setup}
\label{sec:exp:setup}

\subsubsection{Datasets and Tasks}
\label{sec:exp:datasets}

We use six benchmarks%(Table~\ref{tab:datasets})
: two recall-based lists with numeric ground truth, which the model must recall from parametric knowledge ({\em NBA heights} and {\em World population}); large-scale product price sorting ({\em Amazon}); query-driven passage ranking ({\em DL19}); claim--document ranking ({\em SciFact}); and profile-based movie ranking ({\em MovieLens}).

\paragraph{NBA height}
The list contains 200 NBA players with validated heights~\cite{openintro2025}. Following Zhao et al.~\cite{zhao2025accesspaths}, $X$ holds player names only and $c$ requests ascending height with alphabetical tie-breaking, so the LLM must rely on parametric knowledge. The ground-truth order sorts by $(\text{height}, \text{name})$.

\paragraph{World population}
We use the OrderByLLM benchmark~\cite{zhao2025accesspaths}: 200 countries and territories with 2020 populations. As in NBA height, $X$ holds names only, so the LLM must rely on parametric knowledge, and $c$ requests ascending population with alphabetical tie-breaking. The ground-truth order sorts by $(\text{population}, \text{name})$, and quality is measured by Kendall's~$\tau$ over the full list.

\paragraph{Amazon price ranking}
We sample $n{=}1{,}000{,}000$ priced products from the Amazon Reviews 2023 product metadata~\cite{hou2024bridging}, keeping one row per parent ASIN (Amazon's identifier for a product and its variants), prices in $[\$0.50,\$10{,}000]$, and titles of at least three words. Each key is a product title shown with its listed price, so comparisons are nearly error-free. We therefore use this benchmark to test whether \textsf{TBS}' bookkeeping and call volume scale to $10^6$ keys, not to measure semantic quality. The criterion~$c$ asks for ascending price, with ties broken alphabetically by title. \textbf{Gold.} The listed price defines the total order (ties broken by title, then ASIN), and we score Kendall's~$\tau$ against it. The full $1$M list is used for the scalability test; smaller runs use fixed random samples of the same corpus (e.g., $n{=}20$K).

\paragraph{DL19 passage ranking}
We adopt the DL19 protocol of Zhao et al.~\cite{zhao2025accesspaths}: 43 TREC 2019 Deep Learning queries, each paired with its BM25 top-$100$ MS~MARCO v1 passages. For each query, we produce a total order of its $100$ passages by relevance to the query text, which serves as the criterion~$c$; keys are passage ids shown with their full text. We report the mean full-list $\mathrm{nDCG}$ over the 43 queries. \textbf{Gold.} As in TREC, human assessors provide graded relevance labels (qrels) from $0$ to $3$, with $3$ the most relevant.

\paragraph{SciFact claim ranking}
We adopt the SciFact ranking protocol of LOTUS~\cite{patel2024semantic} on a fixed slice of the first 50 of the 300 SciFact test claims~\cite{wadden2020fact}, each paired with its top-$100$ abstracts retrieved by E5-base-v2. For each claim, we produce a total order of its $100$ abstracts by relevance to the claim text, which serves as the criterion~$c$; keys are abstract ids shown with their text. We report the mean full-list $\mathrm{nDCG}$ over the 50 claims. \textbf{Gold.} SciFact provides binary relevance labels (qrels, $0$/$1$) indicating whether an abstract is relevant to the claim.

\paragraph{MovieLens personalized ranking}
MovieLens~\cite{harper2016movielens} has also been used to compare pointwise, pairwise, and listwise LLM prompts~\cite{dai2023uncovering}. We fix 63 users with at least 70 ratings and sample $n{=}70$ of each user's rated movies to sort. The criterion~$c$ is a short natural-language user profile; keys are movie ids shown with title and genre. \textbf{Gold.} User ratings ($1$--$5$) serve as graded gains for full-list nDCG, and we report the mean $\mathrm{nDCG}$ over the 63 users.

\subsubsection{Algorithms and Baselines}
\label{sec:exp:baselines}

\paragraph{Proposed method}
We run \textsf{TBS} (Algorithms~\ref{alg:oracle_tbs}--\ref{alg:helpers}) with an LLM in place of the oracle and six-check validation (\S\ref{sec:method:validation}).
Table~\ref{tab:tbs-variants} names the variants we study.
%\paragraph{TBS configuration.}
%The default \textsf{TBS} configuration is given in \S\ref{sec:exp:impl} and Table~\ref{tab:tbs-variants}.
%Named rows in the tables denote the single switch changed from this default (e.g., \textsf{TBS-AllRows}, \textsf{TBS-NoTrans}); combinations (e.g., \textsf{TBS-EarlyStop}, $\theta{=}0.9$) are written explicitly.
%Batch size~$m$ is selected per model and dataset (\S\ref{sec:method:batchsize}).

\begin{table}[!ht]
  \centering
  \caption{\small Named \textsf{TBS} variants.}
  \label{tab:tbs-variants}
  \footnotesize
  \setlength{\tabcolsep}{3pt}
  \begin{tabularx}{\columnwidth}{@{}ll>{\raggedright\arraybackslash}X@{}}
    \toprule
    \textbf{Name} & \textbf{Axis} & \textbf{Setting / change} \\
    \midrule
    \textsf{TBS} & Default & Round~1; recursive $R_0$ Round~2; diagonals then Pack ($\theta{=}1.0$); transitive inference; reachability index \\
    \textsf{TBS-NoReachIndex} & Index & Without the reachability index (heavy full-pair bookkeeping) \\
    \textsf{TBS-AllRows} & Round~2 & Sort every row recursively (Algorithm~\ref{alg:schedule}) \\
    \textsf{TBS-Window} & Round~2 & Windowed sort of $R_0$ (overlapping windows of size ${\le}m$) \\
    \textsf{TBS-EarlyStop} & Round~3 & Stop diagonals when $\rho_t\ge\theta$ ($\theta{=}0.9$ when enabled) \\
    \textsf{TBS-Pack} & Pack & Enable unresolved \textsf{Pack} after the schedule \\
    \textsf{TBS-NoTrans} & Transitivity & Direct edges only: no skip of resolved batches, no cycle rejection (RQ2 ablation) \\
    \bottomrule
  \end{tabularx}
\end{table}

\paragraph{Baselines}
We compare \textsf{TBS} against the ordering paradigms of \S\ref{sec:related:paradigms} (Table~\ref{tab:related-compare}), each instantiated by a prior system or prompt design (listed below). We \emph{reimplement} every baseline and run it on each of our datasets under the \emph{same} LLM, temperature, shuffled input, evaluation keys, and metrics as \textsf{TBS}. %; we do not import published scores. 
Within each comparison table, \textsf{TBS} and all baselines therefore share one LLM (GPT-4o in RQ1; GPT-4.1, GPT-4o, and GPT-4o-mini when three models are reported).

Table~\ref{tab:related-compare} lists further methods that we do not run as baselines, because they target a different problem. LOTUS~\cite{patel2024semantic} and ListK~\cite{listk2026} are semantic top-$K$ operators; the full-sort path of LOTUS is pairwise quicksort, which we include as QS$_1$. LLM-RankFusion~\cite{zeng2024rankfusion} aggregates several rankings, and trained learning-to-rank rerankers~\cite{nogueira2019passage,liu2009learningtorank} require labeled training data. Sliding-window listwise rerankers~\cite{sun2023chatgpt,ma2023listwise} and Setwise prompting~\cite{zhuang2024setwise} can produce a full order, but they are designed for top-$K$ reranking of retrieved candidates. Among listwise schedules, we therefore compare against OrderByLLM merge sort, which targets a complete \texttt{ORDER BY}. Pairwise ranking prompting~\cite{qin2024large} is represented by QS$_1$.

\begin{itemize}[leftmargin=*]\raggedright
  \item \textbf{Pairwise quicksort} (OrderByLLM QS$_1$~\cite{zhao2025accesspaths}): quicksort with an LLM pairwise comparator~\cite{dai2023uncovering,qin2024large}. Each comparison is one LLM call that orders a single pair, so the sort issues $O(n\log n)$ calls.
  \item \textbf{Pointwise scoring} ($n$ calls): one prompt per key asks the LLM for a numeric score (e.g., predicted height or relevance); replies are parsed as floats, and keys are sorted by score in the direction required by~$c$~\cite{dai2023uncovering,zhao2025accesspaths}. For example, if the LLM returns $2.01$, $1.96$, and $2.11$ for $\textit{PlayerA}$, $\textit{PlayerB}$, and $\textit{PlayerC}$, ascending-height order is $\textit{PlayerB}\prec\textit{PlayerA}\prec\textit{PlayerC}$.
  \item \textbf{OrderByLLM listwise merge sort}~\cite{zhao2025accesspaths}: batched external merge sort that sorts contiguous runs of $m$ keys and then merges runs with further listwise calls. Unlike \textsf{TBS}, it does not reuse relations revealed by earlier calls. We run it at the same batch size~$m$ as \textsf{TBS} in each experiment.
\end{itemize}

\subsubsection{Models}
\label{sec:exp:models}

%\FloatBarrier
We evaluate \textsf{TBS} with four LLMs: %(Table~\ref{tab:models}).
GPT-4.1, GPT-4o, and GPT-4o-mini. %are the baseline comparators (Table~\ref{tab:pairwise-baselines-v2} and the shared-schedule RQs).
GPT-5-nano is used only for \emph{Amazon price ranking} \footnote{\scriptsize Because prices are shown in the prompt, the task requires no factual recall and tests only whether \textsf{TBS}'s bookkeeping and call volume scale to $10^{6}$ keys; a small model suffices and keeps the million-key run affordable}. %: the batch-size sweep at $n{=}200$, the RQ2 ablations at $n{=}200$--$1{,}000$ (Tables~\ref{tab:reachindex-baselines}, \ref{tab:theta-sweep}, and~\ref{tab:round2-rows}), and the scaling runs at $n{=}20$K and $n{=}1$M (Table~\ref{tab:amazon-scaling}).
All calls use temperature~0, except GPT-5-nano, whose API accepts only the default temperature.

%\begin{table}[!ht]
%  \centering
%  \caption{LLMs used in our experiments.}
%  \label{tab:models}
%  \footnotesize
%  \setlength{\tabcolsep}{4pt}
%  \begin{tabular}{@{}lll@{}}
%    \toprule
%    \textbf{Model} & \textbf{API model id} & \textbf{Used in} \\
%    \midrule
%    GPT-4.1        & \texttt{gpt-4.1}       & Baselines \\
%    GPT-4o         & \texttt{gpt-4o}  %      & Baselines \\
%    GPT-4o-mini    & \texttt{gpt-4o-mini}   & Baselines \\
%    GPT-5-nano     & \texttt{gpt-5-nano}    & Amazon \\
%    \bottomrule
%  \end{tabular}
%\end{table}

\subsubsection{Metrics}
\label{sec:exp:metrics}

\paragraph{Quality}
For {\em NBA}, {\em World population}, and {\em Amazon} (each is a single list of $n$ keys), we report Kendall's~$\tau_a$ between the predicted ranking $\pi$ and the ground-truth ranking $\pi^\star$:
\[
\tau_a(\pi,\pi^\star) = \frac{C-D}{\binom{n}{2}},
\]
where $C$ (resp.\ $D$) is the number of key pairs whose relative order agrees (resp.\ disagrees) under $\pi$ and $\pi^\star$, and $\tau_a{=}1$ indicates a perfect match. Since gold ties are broken by name or title, both orders are total and no tie correction is needed.

For {\em DL19}, {\em SciFact}, and {\em MovieLens}, we report the mean of per-instance full-list normalized discounted cumulative gain (nDCG). Given graded gains $g_i$ ({\em DL19}: TREC $0$--$3$; {\em SciFact}: binary $0$/$1$; {\em MovieLens}: ratings $1$--$5$) and predicted ranks $r_i$,
\[
\mathrm{DCG} = \sum_{i=1}^{n} \frac{g_i}{\log_2(r_i+1)},
\qquad
\mathrm{nDCG} = \frac{\mathrm{DCG}}{\mathrm{IDCG}},
\]
where $\mathrm{IDCG}$ is the DCG of the ideal ranking by~$g_i$. We average nDCG over the instances of each benchmark: $\mathrm{nDCG}_Q$ over the $43$ {\em DL19} queries, $\mathrm{nDCG}_C$ over the $50$ {\em SciFact} claims, and $\mathrm{nDCG}_U$ over the $63$ {\em MovieLens} users.

\begin{table}[t]
  \centering
  \caption{\small RQ1: QS$_1$, pointwise, merge, and \textsf{TBS} (GPT-4o, $m{=}10$). TBS uses the best Round~2 setting per dataset. {\em Amazon} uses $n{=}1{,}000$.}
  \label{tab:pairwise-baselines-v2}
  \label{tab:pointwise-baselines}
  \label{tab:merge-baselines}
  \label{tab:rq1-gpt4o}
  \small
  %\setlength{\tabcolsep}{3pt}
  %\resizebox{\columnwidth}{!}{%
  \begin{tabular}{@{}l l rrrr@{}}
    \toprule
    \textbf{Dataset} & \textbf{Metric} & \textbf{QS$_1$} & \textbf{PW} & \textbf{Merge} & \textbf{TBS} \\
    \midrule
    \multirow{4}{*}{\em NBA}
      & $\tau$ $\uparrow$   & \textbf{0.943} & 0.938 & 0.879 & 0.876 \\
      & \# calls $\downarrow$   & 1{,}483 & 200 & 148 & \textbf{97} \\
      & Tokens (M) $\downarrow$ & 0.49 & \textbf{0.05} & 0.09 & 0.06 \\
      & Cost (\$) $\downarrow$  & 3.07 & \textbf{0.34} & 0.64 & 0.43 \\
    \midrule
    \multirow{4}{*}{\em Population}
      & $\tau$  $\uparrow$    & 0.976 & \textbf{0.986} & 0.933 & 0.944 \\
      & \# calls $\downarrow$   & 1{,}421 & 200 & 152 & \textbf{121} \\
      & Tokens (M) $\downarrow$  & 0.37 & \textbf{0.04} & 0.08 & 0.07 \\
      & Cost (\$)  $\downarrow$ & 2.01 & \textbf{0.21} & 0.58 & 0.45 \\
    \midrule
    \multirow{4}{*}{\emph{Amazon}}
      & $\tau\uparrow$          & 1.000 & \textbf{1.000} & 0.999 & 0.972 \\
      & \# calls$\downarrow$    & 10{,}351 & 1{,}000 & 1{,}051 & \textbf{643} \\
      & Tokens (M)$\downarrow$  & 3.53 & \textbf{0.25} & 0.90 & 0.52 \\
      & Cost (\$)$\downarrow$   & 16.92 & \textbf{1.21} & 5.10 & 2.91 \\
    \midrule
    \multirow{4}{*}{\em DL19}
      & $\mathrm{nDCG}_Q$ $\uparrow$  & 0.907 & 0.855 & 0.916 & \textbf{0.921} \\
      & \# calls $\downarrow$    & 30{,}913 & 4{,}300 & 2{,}347 & \textbf{1{,}790} \\
      & Tokens (M) $\downarrow$  & 15.59 & \textbf{1.62} & 3.33 & 2.24 \\
      & Cost (\$)  $\downarrow$ & 79.21 & \textbf{6.28} & 15.43 & 10.49 \\
    \midrule
    \multirow{4}{*}{\em SciFact}
      & $\mathrm{nDCG}_C$ $\uparrow$ & 0.849 & 0.691 & 0.850 & \textbf{0.854} \\
      & \# calls $\downarrow$   & 44{,}047 & 5{,}000 & 2{,}511 & \textbf{2{,}200} \\
      & Tokens (M) $\downarrow$ & 42.11 & \textbf{3.07} & 9.22 & 6.98 \\
      & Cost (\$)  $\downarrow$ & 169.59 & \textbf{10.83} & 33.14 & 25.48 \\
    \midrule
    \multirow{4}{*}{\em MovieLens}
      & $\mathrm{nDCG}_U$ $\uparrow$ & \textbf{0.887} & 0.870 & 0.866 & 0.879 \\
      & \# calls  $\downarrow$  & 24{,}424 & 4{,}410 & 2{,}141 & \textbf{1{,}832} \\
      & Tokens (M) $\downarrow$ & 7.76 & 2.34 & 2.26 & \textbf{0.87} \\
      & Cost (\$)  $\downarrow$ & 20.03 & 11.35 & 13.83 & \textbf{2.61} \\

    \bottomrule
  \end{tabular}%
  %}
  \vspace{-3mm}
\end{table}

\begin{table*}[!t]
\centering
\caption{\small RQ2 ablations: $Q$ (quality) denotes $\tau$ on {\em NBA}/{\em Population}/{\em Amazon} and mean full-list nDCG on {\em DL19}/{\em SciFact}/{\em MovieLens}; \# is LLM calls; Tokens (Tok) in M. (a)~NoTrans vs.\ TBS (GPT-4o-mini). (b)~ReachIndex (RI) vs.\ explicit closure on {\em Amazon} (GPT-5-nano, $m{=}40$); $\tau$/\#/Tok/\$ match for both backends; peak memory (mem.) in MB; 20K/1M use only RI with $\theta{=}0.9$.
(c)~$\theta{=}0.90$ vs.\ full packing (GPT-4o; {\em Amazon}$_{1\mathrm{k}}$: GPT-5-nano).
(d)~Round~2 $R_0$/all-rows $\times$ recursive (Rec.)/window (Win.).}
\label{tab:rq2-ablations}
\label{tab:no-trans}\label{tab:notrans-baselines}\label{tab:notrans}
\label{tab:reachindex-baselines}\label{tab:amazon-scaling}
\label{tab:theta-sweep}
\label{tab:round2-rows}
\vspace{-1mm}
\scriptsize
\setlength{\tabcolsep}{1.8pt}
\renewcommand{\arraystretch}{0.95}
\begin{minipage}[t]{0.495\textwidth}
\centering
\textbf{(a) Transitivity}\\[1pt]
\begin{tabular}{@{}l rrrr rrrr@{}}
\toprule
& \multicolumn{4}{c}{\textbf{NoTrans}} & \multicolumn{4}{c}{\textbf{TBS}} \\
\cmidrule(lr){2-5}\cmidrule(l){6-9}
\textbf{Dataset} & $Q{\uparrow}$ & \#${\downarrow}$ & Tok${\downarrow}$ & \$\,$\downarrow$ & $Q{\uparrow}$ & \#${\downarrow}$ & Tok${\downarrow}$ & \$\,$\downarrow$ \\
\midrule
{\em NBA} ($m{=}4$)
  & 0.354 & 3{,}436 & 1.18 & 0.41
  & \textbf{0.395} & \textbf{306} & \textbf{0.09} & \textbf{0.02} \\
{\em Pop.} ($m{=}6$)
  & 0.946 & 1{,}532 & 0.64 & 0.25
  & \textbf{0.952} & \textbf{202} & \textbf{0.08} & \textbf{0.03} \\
{\em DL19} ($m{=}5$)
  & \textbf{0.915} & 24{,}469 & 19.03 & 4.71
  & 0.900 & \textbf{4{,}522} & \textbf{3.40} & \textbf{0.85} \\
{\em SciFact} ($m{=}4$)
  & 0.841 & 44{,}545 & 70.56 & 14.72
  & \textbf{0.871} & \textbf{7{,}392} & \textbf{11.48} & \textbf{2.42} \\
{\em ML} ($m{=}10$)
  & 0.848 & 5{,}859 & 3.06 & 0.59
  & \textbf{0.878} & \textbf{1{,}784} & \textbf{0.86} & \textbf{0.16} \\
\bottomrule
\end{tabular}
\end{minipage}\hfill
\begin{minipage}[t]{0.495\textwidth}
\centering
\textbf{(b) Reachability ({\em Amazon})}\\[1pt]
\begin{tabular}{@{}r rrrr rr rr@{}}
\toprule
& \multicolumn{4}{c}{\textbf{Shared}} & \multicolumn{2}{c}{\textbf{Time (s)}${\downarrow}$} & \multicolumn{2}{c}{\textbf{Peak MB}${\downarrow}$} \\
\cmidrule(lr){2-5}\cmidrule(lr){6-7}\cmidrule(l){8-9}
$n$ & $\tau{\uparrow}$ & \#${\downarrow}$ & Tok${\downarrow}$ & \$\,$\downarrow$ & NoRI & RI & NoRI & RI \\
\midrule
200
  & 0.9999 & 57 & 0.13 & 0.04
  & 2.19 & \textbf{0.05} & 2.5 & \textbf{0.1} \\
500
  & 0.9997 & 94 & 0.39 & 0.13
  & 34.8 & \textbf{0.40} & 17.0 & \textbf{0.6} \\
1{,}000
  & 0.999 & 171 & 0.96 & 0.33
  & 495.1 & \textbf{0.41} & 49.8 & \textbf{0.6} \\
\midrule
20K & 0.946 & 1{,}421 & 11.18 & 3.78 & -- & -- & -- & -- \\
1M  & 0.923 & 91{,}383 & 829.45 & 286.37 & -- & -- & -- & -- \\
\bottomrule
\end{tabular}
\end{minipage}

\vspace{1.5mm}
\begin{minipage}[t]{0.495\textwidth}
\centering
\textbf{(c) Early stop ($\theta{=}0.90$ vs.\ full)}\\[1pt]
\begin{tabular}{@{}l rrrr rrrr@{}}
\toprule
& \multicolumn{4}{c}{$\theta{=}0.90$} & \multicolumn{4}{c}{\textbf{Full}} \\
\cmidrule(lr){2-5}\cmidrule(l){6-9}
\textbf{Dataset} & $Q{\uparrow}$ & \#${\downarrow}$ & Tok${\downarrow}$ & \$\,$\downarrow$ & $Q{\uparrow}$ & \#${\downarrow}$ & Tok${\downarrow}$ & \$\,$\downarrow$ \\
\midrule
{\em NBA}
  & 0.851 & \textbf{64} & 0.04 & \textbf{0.28}
  & \textbf{0.876} & 97 & 0.06 & 0.43 \\
{\em Population}
  & 0.867 & \textbf{67} & 0.04 & \textbf{0.24}
  & \textbf{0.944} & 121 & 0.07 & 0.45 \\
{\em Amazon}$_{1\mathrm{k}}$
  & 0.949 & \textbf{102} & 0.50 & \textbf{0.17}
  & \textbf{0.9997} & 171 & 0.96 & 0.33 \\
{\em DL19}
  & 0.885 & \textbf{1{,}279} & 1.64 & \textbf{7.71}
  & \textbf{0.921} & 1{,}790 & 2.24 & 10.49 \\
{\em SciFact}
  & 0.528 & \textbf{1{,}469} & 4.64 & \textbf{17.01}
  & \textbf{0.854} & 2{,}200 & 6.98 & 25.48 \\
{\em MovieLens}
  & 0.876 & \textbf{1{,}320} & 0.63 & \textbf{1.90}
  & \textbf{0.879} & 1{,}832 & 0.87 & 2.61 \\
\bottomrule
\end{tabular}
\end{minipage}\hfill
\begin{minipage}[t]{0.495\textwidth}
\centering
\textbf{(d) Round~2 ($R_0$/all-rows $\times$ Rec./Win.)}\\[1pt]
\begin{tabular}{@{}l rr rr rr rr@{}}
\toprule
& \multicolumn{4}{c}{\textbf{$R_0$ only}} & \multicolumn{4}{c}{\textbf{All rows}} \\
\cmidrule(lr){2-5}\cmidrule(l){6-9}
& \multicolumn{2}{c}{Rec.} & \multicolumn{2}{c}{Win.}
& \multicolumn{2}{c}{Rec.} & \multicolumn{2}{c}{Win.} \\
\cmidrule(lr){2-3}\cmidrule(lr){4-5}\cmidrule(lr){6-7}\cmidrule(l){8-9}
\textbf{Dataset} & $Q{\uparrow}$ & \#${\downarrow}$ & $Q{\uparrow}$ & \#${\downarrow}$ & $Q{\uparrow}$ & \#${\downarrow}$ & $Q{\uparrow}$ & \#${\downarrow}$ \\
\midrule
{\em NBA}
  & \textbf{0.876} & 97 & 0.852 & \textbf{94}
  & 0.841 & 108 & 0.861 & 103 \\
{\em Population}
  & 0.932 & 101 & 0.926 & \textbf{98}
  & \textbf{0.944} & 121 & 0.927 & 109 \\
{\em DL19}
  & \textbf{0.921} & \textbf{1{,}790} & \textbf{0.921} & \textbf{1{,}790}
  & 0.911 & 1{,}847 & 0.911 & 1{,}847 \\
{\em SciFact}
  & 0.839 & \textbf{2{,}102} & 0.839 & \textbf{2{,}102}
  & \textbf{0.854} & 2{,}200 & \textbf{0.854} & 2{,}200 \\
{\em MovieLens}
  & \textbf{0.879} & \textbf{1{,}832} & \textbf{0.879} & \textbf{1{,}832}
  & 0.870 & 2{,}085 & 0.870 & 2{,}085 \\
{\em Amazon}$_{1\mathrm{k}}$
  & \textbf{0.999} & 171 & \textbf{0.999} & 171
  & \textbf{0.999} & \textbf{165} & \textbf{0.999} & \textbf{165} \\
\bottomrule
\end{tabular}
\end{minipage}
\vspace{-5mm}
\end{table*}

\paragraph{Cost}
We report three cost metrics: (i)~the number of LLM calls per sort (per query, claim, or user on multi-instance datasets), excluding the optional robustness check; (ii)~total tokens (prompt plus completion, as returned by the provider); and (iii)~estimated USD cost at the model's published list rates. Calls and dollar cost capture different costs and are reported separately: calls are limited by API rate limits and per-request overhead, whereas dollar cost depends on tokens, so a method with many short prompts, such as pointwise scoring, can issue more calls than \textsf{TBS} yet cost less.

\subsubsection{Implementation}
\label{sec:exp:impl}
\label{sec:exp:system}

\textsf{TBS} is implemented in Python~3. All batch calls share one listwise prompt layout with structured output, following the protocol of Zhao et al.~\cite{zhao2025accesspaths}; tasks differ only in how keys are rendered (names, passage or claim text, movie title and genre, or {\em Amazon} title with price). Each prompt also states the alphabetical tie-break of its gold order.

%\paragraph{Default \textsf{TBS}.}
%Unless a named variant or ablation says otherwise, \textsf{TBS} means: contiguous Round~1; direct-edge reachability index (\textsf{TBS-ReachIndex}, \S\ref{sec:method:dag}); recursive $R_0$ Round~2; diagonals then unresolved \textsf{Pack} with $\theta{=}1.0$; validation and cycle-inducing rejection enabled. The default comparator for baseline tables is GPT-4o (Table~\ref{tab:models}); GPT-4.1 and GPT-4o-mini appear in multi-model RQs, and GPT-5-nano only on \emph{Amazon}.

\paragraph{Batch size}
We select $m$ per model and dataset from the candidate set $\mathcal{M}=\{4,\dots,14\}$ using the procedure of \S\ref{sec:method:batchsize}. For GPT-4o the sweep selects $m{=}10$ on every dataset, so all RQ1 comparisons use $m{=}10$; in RQ3, each model runs at its own~$m$, since smaller models order fewer items reliably per call. For {\em Amazon}, which targets million-item lists, we sweep larger candidates $\mathcal{M}_{\text{Amazon}}=\{10,20,40,75\}$ at $n{=}200$ and choose $m{=}40$: $m{=}20$ and $m{=}40$ tie on quality and cost, and the larger value halves the number of Round-1 columns at $n{=}10^6$. When early stopping is enabled, we use $\theta{=}0.9$ with a Round~3 stall limit of~2; all other runs use $\theta{=}1.0$.

\paragraph{Baselines}
All baselines are reimplemented under the same LLM, temperature, shuffled input, and evaluation keys as TBS, with merge sort run at the same batch size~$m$.

\subsection{Results}
\label{sec:exp:results}

Due to space constraints, we report results for RQ1 and RQ2 here; results for RQ3--RQ5 appear in the Appendix.

\subsubsection{RQ1: Quality--Cost Trade-off}
\label{sec:exp:rq1}
\label{sec:exp:nba}

Following the protocol of \S\ref{sec:exp:baselines}, we compare \textsf{TBS} with pairwise, pointwise, and listwise baselines under a single LLM (GPT-4o) across all six datasets, using the best Round~2 setting per dataset from Table~\ref{tab:round2-rows}. Table~\ref{tab:pairwise-baselines-v2} reports ranking quality and execution cost. 
\paragraph{Pairwise QS$_1$} Compared with pairwise quicksort, \textsf{TBS} consistently achieves substantial cost reductions in API calls, tokens, and monetary expense. The gap arises because \textsf{TBS} reuses preferences through transitive inference, whereas QS$_1$ resolves comparisons independently. In terms of quality, the picture is mixed. QS$_1$ remains stronger on the recall-based sorting tasks, where the model can often answer individual comparisons directly from memorized facts. However, \textsf{TBS} exceeds QS$_1$ on {\em DL19} and {\em SciFact}, and trails slightly on {\em MovieLens} ($0.879$ vs.\ $0.887$), suggesting that aggregating evidence across listwise batches is often more effective when ranking requires nuanced comparative judgments rather than factual recall. \paragraph{Pointwise scoring} Pointwise scoring is competitive on {\em NBA}, {\em population}, and {\em Amazon} ranking because each item has an underlying absolute attribute that can be estimated independently. In such settings, assigning a score to each key is both simple and accurate. This advantage largely disappears on relevance and preference ranking tasks, where quality depends on relative comparisons rather than absolute values. There, \textsf{TBS} consistently outperforms pointwise scoring, indicating that listwise reasoning captures contextual relationships that independent scoring prompts fail to model reliably. \paragraph{Listwise merge sort} The comparison with merge sort isolates the effect of transitive reuse, since both methods rely on listwise prompting. Across datasets, \textsf{TBS} achieves comparable or higher ranking quality while requiring fewer LLM interactions and lower overall cost. The advantage stems from maintaining a preference graph and avoiding batches whose ordering is already implied by earlier answers, whereas merge sort repeatedly revisits previously established relations as part of its fixed merge schedule. \paragraph{Takeaway} The results reveal a clear task-dependent trade-off. Pointwise scoring is the strongest option when the ranking criterion corresponds to an independently estimable numeric attribute, such as height or population. In contrast, \textsf{TBS} is consistently more effective for semantic relevance and preference ranking, where relative comparisons matter more than absolute scores. Relative to pairwise and listwise baselines, \textsf{TBS} achieves these quality levels with substantially fewer LLM calls by reusing previously acquired preferences rather than repeatedly querying them. Overall, the results support the central premise of \textsf{TBS}: transitivity-aware preference reuse yields significant efficiency gains while preserving, and often improving, ranking quality on semantic ordering tasks.

\subsubsection{RQ2: Contribution of Optimizations}
\label{sec:exp:transitivity}
\label{sec:exp:rq2}
We ablate one switch at a time from the default (\S\ref{sec:exp:impl}).
Unless noted, runs use GPT-4o, $m{=}10$, and the prompts/inputs of Table~\ref{tab:pairwise-baselines-v2}.
Table~\ref{tab:rq2-ablations} summarizes all four ablations.

\spara{Transitivity.}
\textsf{TBS-NoTrans} skips no batch and packs until every pair is observed (Table~\ref{tab:rq2-ablations}a; GPT-4o-mini at its RQ3~$m$).
Transitivity cuts calls by $3$--$11\times$ on every dataset.
Quality is mixed: NoTrans leads only on {\em DL19}; TBS leads on the other four, so the main win is still cost under a noisy comparator.

\spara{Reachability index.}
On \emph{Amazon} (GPT-5-nano, $m{=}40$), both backends replay the same answers, so $\tau$, calls, tokens, and \$ match (Table~\ref{tab:rq2-ablations}b).
ReachIndex is $45$--$1{,}200\times$ faster with $25$--$83\times$ less memory, enabling $n{=}20$K ($\tau{=}0.946$) and $n{=}1$M ($\tau{=}0.923$, \$$286$; $\theta{=}0.9$).

\spara{Early stopping.}
Stopping Round~3 once $\rho_t\ge\theta$ saves $28$--$45\%$ of calls at $\theta{=}0.90$ (Table~\ref{tab:rq2-ablations}c).
{\em NBA}/{\em MovieLens} stay near full quality; {\em SciFact} drops sharply because one relevant abstract per claim often remains unresolved.

\spara{Round~2.}
Recursive $R_0$-only is best or tied on {\em NBA}, {\em DL19}, {\em MovieLens}, and {\em Amazon} at the fewest or near-fewest calls (Table~\ref{tab:rq2-ablations}d); windows cut a few more calls on {\em NBA}/{\em population}.
All-rows helps only {\em population} ($+0.012$) and {\em SciFact} ($+0.015$), at $5$--$20\%$ more calls.
Token and dollar costs track the call counts.

\spara{Takeaway.}
Transitive reuse is the main cost lever; the reachability index enables $10^6$ keys; recursive $R_0$ is a sound default; $\theta{<}1$ is reasonable when relevance is spread over many keys.

\section{Conclusions}
\label{sec:conclusion}

We presented Transitive Batched Sort (\textsf{TBS}), a listwise \textsc{order by} operator that records preferences in an acyclic graph and skips pairs already implied by transitivity. A direct-edge reachability index keeps memory linear and lets \textsf{TBS} scale to $10^6$ keys on {\em Amazon}.
On the six datasets with GPT-4o baselines, \textsf{TBS} uses $12$--$20\times$ fewer calls than pairwise quicksort and $1.1$--$1.6\times$ fewer than listwise merge sort, with higher quality than merge on four datasets ($0.003$ lower on {\em NBA} and $0.027$ lower on {\em Amazon}) and the highest quality across all methods on {\em DL19} and {\em SciFact}.
\textsf{TBS} is best for budget-constrained ranking without an absolute numeric scale; while pointwise scoring remains stronger on {\em NBA}, {\em Population}, and {\em Amazon}, and QS$_1$ on {\em NBA}, {\em Population}, {\em Amazon}, and {\em MovieLens}.
\enlargethispage{3\baselineskip}
Future work includes transitivity-aware top-$K$ selection, cost-based access-path selection, and setting optimal $m$ and $\theta$ without a sweep.

\bibliographystyle{IEEEtran}
\bibliography{references}

\newpage 

\appendices
\raggedbottom
\setlength{\textfloatsep}{6pt plus 1pt minus 1pt}
\setlength{\floatsep}{5pt plus 1pt minus 1pt}
\setlength{\intextsep}{5pt plus 1pt minus 1pt}
\setlength{\abovecaptionskip}{2pt}
\setlength{\belowcaptionskip}{6pt}

\section{Notations}
Table~\ref{tab:notations} lists notation used in the paper.
\begin{center}
  \captionof{table}{\small Frequently used notations.}
  \label{tab:notations}
  \scriptsize
  \setlength{\tabcolsep}{3pt}
  \renewcommand{\arraystretch}{1.05}
  \begin{tabularx}{\columnwidth}{@{}l>{\raggedright\arraybackslash}X@{}}
    \toprule
    \textbf{Notation} & \textbf{Meaning} \\
    \midrule
    $n$ & number of keys (items) to be sorted \\
    $X=\{x_1,\ldots,x_n\}$ & input key list submitted to the sorting operator \\
    $c$ & natural-language ordering criterion (prompt) \\
    $m$ & listwise batch size ($m\ge 2$); knee sweep tries $m\in\mathcal{M}$, then fixes one per model \\
    $\mathcal{M}$ & candidate batch sizes ($\{4,\ldots,14\}$; $\{10,20,40,75\}$ on Amazon) \\
    $B$ & subset of keys in one listwise LLM call \\
    $\sigma(B)=(x_1,\ldots,x_m)$ & permutation of $B$ returned by the LLM \\
    $a \succ b$ & strict preference: $a$ ranks above $b$ under~$c$ \\
    $\mathcal{R}$ & maintained ordered pairs $(a,b)$ with $a \succ b$ \\
    $G=(V,E)$ & relation graph, $V=X$, edge $u\!\to\!v$ iff $(u,v)\in\mathcal{R}$ \\
    $H=(X,E_d)$ & sparse DAG of direct edges ($G$ is its transitive closure) \\
    $\pi$ & final total order over $X$ (lexicographic tie-break) \\
    \bottomrule
  \end{tabularx}
\end{center}

\section{Additional Related Work}
\label{sec:app:related}

\spara{Batched comparators.}
Ordering $k$ items per call is classical: Beigel and Gill bound $k$-sorter complexity~\cite{beigel1990ksorter}; Columnsort~\cite{leighton1985columnsort} and Shearsort~\cite{scherson1986shearsort} alternate column and row phases.
\textsf{TBS} reuses that grid for Rounds~1--3, but unlike those perfect-comparator schedules it skips batches already implied by earlier answers and rejects contradictions.

\spara{Crowdsourced sorting.}
Crowd DB operators also pay per comparison and tolerate noise~\cite{marcus2011humanpowered},~\cite{guo2012sowhowon,venetis2012max,davidson2013topk,chen2013pairwise}; several save questions via transitivity.
\textsf{TBS} applies the same idea to listwise LLM calls, with three differences: each call yields up to $\binom{m}{2}$ preferences, the schedule is the batch grid, and the graph must stay cheap at large~$n$ (addressed by the reachability index, \S\ref{sec:method:dag}).

\spara{Noisy sorting.}
Theory bounds comparisons under random errors with and without resampling~\cite{feige1994noisy,braverman2008noisy} and studies active ranking under geometric assumptions~\cite{jamieson2011active}.
\textsf{TBS} assumes no error model: it only keeps $G$ acyclic and measures residual disagreement empirically (\S\ref{sec:method:errors}).

\FloatBarrier
\section{Additional Experimental Results}
\label{sec:app:additional-results}

\subsection{Quantifying Residual LLM Inconsistency}
\label{sec:method:errors}

Format validation (\S\ref{sec:method:validation}) ensures that a reply is a permutation of its batch, not that preferences are correct. Two residual inconsistencies remain, summarized in Table~\ref{tab:rq2-inconsistency}.

\begin{center}
\captionof{table}{\small \textsf{TBS} inconsistency (GPT-4o).
$\varepsilon_{\mathrm{reject}}$, $\eta$, Malformed: Round~2 $R_0$ audit ({\em NBA}/{\em Pop.}/{\em Amazon}: one instance; {\em DL19}/{\em SciFact}/{\em MovieLens}: first 20).
Pack calls: full sort to $u{=}0$ with all-rows Round~2; {\em NBA}/{\em Pop.}/{\em Amazon}: one instance; {\em DL19}/{\em SciFact}/{\em MovieLens}: first 20.}
\label{tab:rq2-inconsistency}
\scriptsize
\setlength{\tabcolsep}{2.6pt}
\renewcommand{\arraystretch}{1.08}
\begin{tabularx}{\columnwidth}{@{}l*{6}{>{\raggedleft\arraybackslash}X}@{}}
\toprule
\textbf{Metric} & \textbf{\em NBA} & \textbf{\em Pop.} & \textbf{\em DL19} & \textbf{\em SciFact} & \textbf{\em ML} & \textbf{\em Amz.} \\
\midrule
$\varepsilon_{\mathrm{reject}}$ $\downarrow$ & 0.067 & 0.065 & 0.102 & 0.113 & 0.091 & 0.041 \\
$\eta$ $\downarrow$ & 0.038 & 0.063 & 0.115 & 0.037 & 0.178 & 0.030 \\
Pack calls (\%) $\downarrow$ & 12 & 11 & 9 & 11 & 5 & 16 \\
Malformed (\%) $\downarrow$ & 0.0 & 0.0 & 0.4 & 1.0 & 0.0 & 0.0 \\
\bottomrule
\end{tabularx}
\end{center}

\spara{Detectable contradictions.}
An edge that would cycle $G$ is rejected at merge. With $D$ accepted and $R_{\mathrm{rej}}$ rejected direct edges, $\varepsilon_{\mathrm{reject}}=R_{\mathrm{rej}}/(D+R_{\mathrm{rej}})$ is a merge-time conflict rate, not a lower bound on LLM mistakes.

\spara{Undetectable semantic noise.}
Edges consistent with $G$ may still be wrong under~$c$. We re-ask stored direct edges pairwise; if $C$ is the fraction whose direction agrees with the stored edge, $\eta = 1 - C$. The two rates use different denominators and are not combined.

Table~\ref{tab:rq2-inconsistency} separates format errors from preference noise.
When a listwise reply proposes an edge that would cycle the preference graph~$G$, \textsf{TBS} rejects that edge at merge time: across datasets, $\varepsilon_{\mathrm{reject}}$ ranges from $0.041$ on {\em Amazon} to $0.113$ on {\em SciFact} (about $4$--$11\%$ of candidate direct edges).
This rate indicates conflicts with already-accepted structure, not all LLM mistakes. An early wrong edge can close off later correct ones, so a later true preference may be discarded even though the model answered correctly on that batch.
That effect helps explain why disabling transitivity (\textsf{TBS-NoTrans}) improves quality only on {\em DL19} (Table~\ref{tab:no-trans}).

Edges that do enter~$G$ can still be semantically wrong.
Re-asking a sample of stored direct edges as fresh pairwise comparisons, the disagreement rate is $\eta{=}0.030$--$0.178$: lowest on {\em Amazon} (prices shown in the prompt) and {\em SciFact}, and highest on {\em MovieLens}, where preference judgments are more subjective.

Separately, under all-rows Round~2 with packing run to completion ($u{=}0$), unresolved \textsf{Pack} accounts for $5$--$16\%$ of all LLM calls (lowest on {\em MovieLens}, highest on {\em Amazon}).

Malformed replies, i.e., outputs that are not a clean permutation of the batch, are at most $1\%$ ({\em SciFact}) and $0\%$ on four datasets; every such reply is repaired by retry before any edge is merged.

\subsection{RQ3: Robustness across Models and Tasks}
\label{sec:exp:rq3}

We compare GPT-4.1, GPT-4o, and GPT-4o-mini under the default schedule (\S\ref{sec:exp:impl}) at each model's selected~$m$ (Table~\ref{tab:rq3-models}; sweeps in Figure~\ref{fig:m-sweep-datasets}).
Mini needs smaller~$m$ (e.g., {\em population} $\tau{=}0.952$ at $m{=}6$ vs.\ $0.564$ at $m{=}10$).
On total orders, GPT-4o leads {\em NBA} ($0.876$/$97$ calls) over GPT-4.1 ($0.817$/$110$), while mini collapses ($0.208$/$306$); population is easy for all ($\tau{\ge}0.93$; GPT-4.1 highest at $0.972$/$87$).
On relevance/preference, quality is close ({\em MovieLens} $0.878$--$0.879$; {\em SciFact} $0.839$--$0.871$, mini best; {\em DL19} $0.900$--$0.921$, GPT-4o best), but mini is cheapest in dollar cost despite often more calls.

\spara{Takeaway.}
Model choice matters most when the ranking requires recalling factual knowledge from model parameters (e.g., {\em NBA height} and {\em population} ranking). Elsewhere, ranking quality is relatively similar across models, and dollar cost is influenced more by token pricing than by call count.

\subsection{RQ4: Batch-size Sensitivity}
\label{sec:exp:rq4}
\label{sec:exp:batchsize}

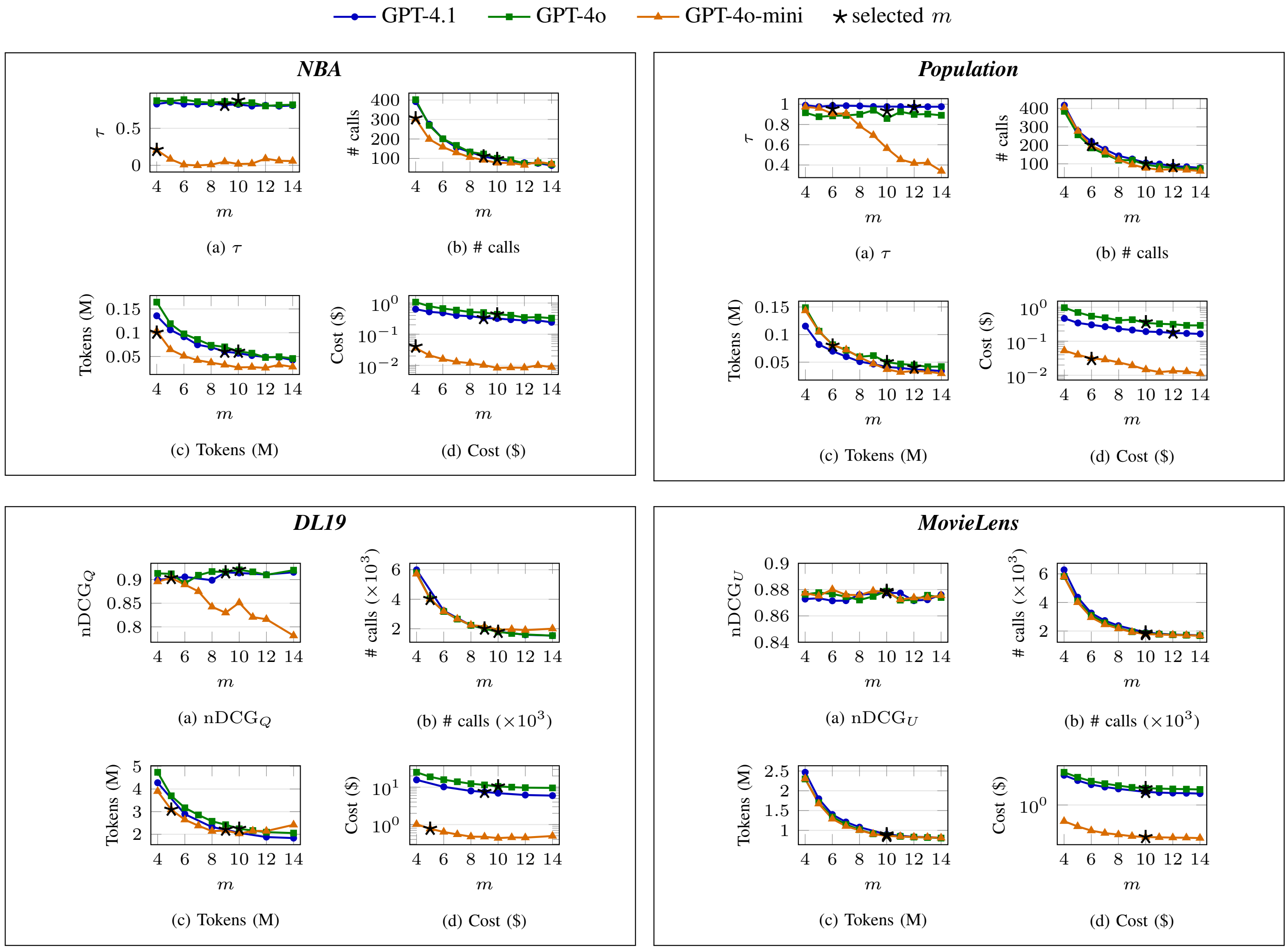
\begin{figure*}[p]
\centering
\pgfplotsset{
  sweep/.style={width=3.70cm, height=2.70cm, xmin=3.5, xmax=14.5, xtick={4,6,...,14}, xlabel={$m$},
    tick label style={font=\scriptsize}, label style={font=\scriptsize},
    title style={at={(0.5,-0.95)}, anchor=north, font=\scriptsize},
    ylabel style={yshift=-0.5ex},
    ymajorgrids, grid style={gray!20}, scaled y ticks=false,
    yticklabel style={/pgf/number format/fixed, /pgf/number format/precision=3}},
  m41/.style={blue!75!black, thick, mark=*, mark size=1.0pt},
  m4o/.style={green!50!black, thick, mark=square*, mark size=0.95pt},
  mmini/.style={orange!85!black, thick, mark=triangle*, mark size=1.3pt},
  sel/.style={only marks, mark=star, mark size=2.9pt, black, thick},
}
% shared legend
\begin{tikzpicture}
\begin{axis}[hide axis, scale only axis, xmin=0, xmax=1, ymin=0, ymax=1, width=0.4cm, height=0.4cm,
  legend style={font=\small, draw=none, legend columns=4, /tikz/every even column/.append style={column sep=10pt}}]
\addlegendimage{m41}\addlegendentry{GPT-4.1}
\addlegendimage{m4o}\addlegendentry{GPT-4o}
\addlegendimage{mmini}\addlegendentry{GPT-4o-mini}
\addlegendimage{sel}\addlegendentry{selected $m$}
\end{axis}
\end{tikzpicture}\\[4pt]
\fbox{\begin{minipage}[t]{0.480\textwidth}\centering
{\small\textbf{\em NBA}}\\[4pt]
\begin{tikzpicture}
\begin{groupplot}[group style={group size=2 by 2, horizontal sep=1.55cm, vertical sep=1.75cm}, sweep]
\nextgroupplot[ title={(a) $\tau$}, ylabel={$\tau$}]
\addplot[m41] coordinates {(4,0.8281) (5,0.8538) (6,0.8268) (7,0.8246) (8,0.8323) (9,0.8167) (10,0.8226) (11,0.8008) (12,0.8108) (13,0.7998) (14,0.8081)};
\addplot[m4o] coordinates {(4,0.8735) (5,0.8673) (6,0.887) (7,0.8606) (8,0.8461) (9,0.8618) (10,0.8362) (11,0.8463) (12,0.8024) (13,0.8141) (14,0.8181)};
\addplot[mmini] coordinates {(4,0.206633) (5,0.083819) (6,0.007437) (7,-0.001809) (8,0.008643) (9,0.048442) (10,0.017487) (11,0.022312) (12,0.09005) (13,0.061206) (14,0.057588)};
\addplot[sel] coordinates {(9,0.817) (10,0.876) (4,0.208)};
\nextgroupplot[ title={(b) \# calls}, ylabel={\# calls}]
\addplot[m41] coordinates {(4,392) (5,276) (6,201) (7,156) (8,133) (9,110) (10,97) (11,86) (12,78) (13,75) (14,64)};
\addplot[m4o] coordinates {(4,402) (5,270) (6,201) (7,167) (8,133) (9,123) (10,101) (11,93) (12,74) (13,78) (14,70)};
\addplot[mmini] coordinates {(4,303) (5,199) (6,159) (7,130) (8,106) (9,92) (10,77) (11,77) (12,66) (13,83) (14,72)};
\addplot[sel] coordinates {(9,110) (10,97) (4,306)};
\nextgroupplot[ title={(c) Tokens (M)}, ylabel={Tokens (M)}]
\addplot[m41] coordinates {(4,0.1356) (5,0.1061) (6,0.0915) (7,0.0746) (8,0.0694) (9,0.0603) (10,0.0569) (11,0.052) (12,0.0488) (13,0.0482) (14,0.0423)};
\addplot[m4o] coordinates {(4,0.1644) (5,0.119) (6,0.0976) (7,0.0858) (8,0.0735) (9,0.0702) (10,0.0609) (11,0.0569) (12,0.0479) (13,0.0494) (14,0.0457)};
\addplot[mmini] coordinates {(4,0.103402) (5,0.064647) (6,0.051034) (7,0.042066) (8,0.037004) (9,0.03229) (10,0.026952) (11,0.027573) (12,0.025953) (13,0.032328) (14,0.028176)};
\addplot[sel] coordinates {(9,0.06) (10,0.06) (4,0.1)};
\nextgroupplot[ymode=log, log basis y=10, title={(d) Cost (\$)}, ylabel={Cost (\$)}]
\addplot[m41] coordinates {(4,0.6371) (5,0.5228) (6,0.4839) (7,0.3984) (8,0.3793) (9,0.3335) (10,0.3197) (11,0.2941) (12,0.2762) (13,0.2757) (14,0.2412)};
\addplot[m4o] coordinates {(4,1.0707) (5,0.7897) (6,0.6658) (7,0.5932) (8,0.5163) (9,0.4958) (10,0.4345) (11,0.4062) (12,0.3449) (13,0.3537) (14,0.3269)};
\addplot[mmini] coordinates {(4,0.0361) (5,0.0212) (6,0.0159) (7,0.013) (8,0.0118) (9,0.0102) (10,0.0083) (11,0.0085) (12,0.0084) (13,0.01) (14,0.0088)};
\addplot[sel] coordinates {(9,0.33) (10,0.43) (4,0.04)};
\end{groupplot}
\end{tikzpicture}
\end{minipage}}\hfill
\fbox{\begin{minipage}[t]{0.480\textwidth}\centering
{\small\textbf{\em Population}}\\[4pt]
\begin{tikzpicture}
\begin{groupplot}[group style={group size=2 by 2, horizontal sep=1.55cm, vertical sep=1.75cm}, sweep]
\nextgroupplot[ title={(a) $\tau$}, ylabel={$\tau$}]
\addplot[m41] coordinates {(4,0.9898) (5,0.9755) (6,0.987) (7,0.9849) (8,0.9831) (9,0.9768) (10,0.9767) (11,0.9761) (12,0.972) (13,0.975) (14,0.9757)};
\addplot[m4o] coordinates {(4,0.916) (5,0.8767) (6,0.8851) (7,0.8892) (8,0.9) (9,0.9409) (10,0.8592) (11,0.9257) (12,0.8993) (13,0.9028) (14,0.8907)};
\addplot[mmini] coordinates {(4,0.9733) (5,0.9615) (6,0.9057) (7,0.9089) (8,0.7848) (9,0.6932) (10,0.5643) (11,0.4543) (12,0.4177) (13,0.4223) (14,0.3405)};
\addplot[sel] coordinates {(12,0.972) (10,0.932) (6,0.952)};
\nextgroupplot[ title={(b) \# calls}, ylabel={\# calls}]
\addplot[m41] coordinates {(4,417) (5,278) (6,220) (7,178) (8,143) (9,126) (10,105) (11,99) (12,87) (13,84) (14,78)};
\addplot[m4o] coordinates {(4,384) (5,257) (6,186) (7,152) (8,120) (9,120) (10,95) (11,85) (12,79) (13,73) (14,74)};
\addplot[mmini] coordinates {(4,405) (5,277) (6,197) (7,164) (8,122) (9,95) (10,77) (11,68) (12,70) (13,67) (14,61)};
\addplot[sel] coordinates {(12,87) (10,101) (6,202)};
\nextgroupplot[ title={(c) Tokens (M)}, ylabel={Tokens (M)}]
\addplot[m41] coordinates {(4,0.1153) (5,0.082) (6,0.0699) (7,0.0601) (8,0.051) (9,0.0466) (10,0.0412) (11,0.0396) (12,0.0367) (13,0.0356) (14,0.034)};
\addplot[m4o] coordinates {(4,0.1487) (5,0.1061) (6,0.0816) (7,0.0716) (8,0.0598) (9,0.0619) (10,0.0498) (11,0.0467) (12,0.0449) (13,0.0417) (14,0.042)};
\addplot[mmini] coordinates {(4,0.1435) (5,0.1054) (6,0.0813) (7,0.0725) (8,0.0588) (9,0.0472) (10,0.037) (11,0.0319) (12,0.0341) (13,0.0328) (14,0.0295)};
\addplot[sel] coordinates {(12,0.04) (10,0.05) (6,0.08)};
\nextgroupplot[ymode=log, log basis y=10, title={(d) Cost (\$)}, ylabel={Cost (\$)}]
\addplot[m41] coordinates {(4,0.4689) (5,0.3439) (6,0.3031) (7,0.268) (8,0.2318) (9,0.2134) (10,0.1925) (11,0.1865) (12,0.1753) (13,0.1696) (14,0.1639)};
\addplot[m4o] coordinates {(4,0.9651) (5,0.7005) (6,0.5444) (7,0.4879) (8,0.4114) (9,0.429) (10,0.3446) (11,0.325) (12,0.3146) (13,0.2919) (14,0.2919)};
\addplot[mmini] coordinates {(4,0.0531) (5,0.0399) (6,0.0315) (7,0.0288) (8,0.024) (9,0.0193) (10,0.0147) (11,0.0124) (12,0.0134) (13,0.0129) (14,0.0113)};
\addplot[sel] coordinates {(12,0.18) (10,0.36) (6,0.03)};
\end{groupplot}
\end{tikzpicture}
\end{minipage}}\\[10pt]
\fbox{\begin{minipage}[t]{0.480\textwidth}\centering
{\small\textbf{\em DL19}}\\[4pt]
\begin{tikzpicture}
\begin{groupplot}[group style={group size=2 by 2, horizontal sep=1.55cm, vertical sep=1.75cm}, sweep]
\nextgroupplot[ title={(a) $\mathrm{nDCG}_Q$}, ylabel={$\mathrm{nDCG}_Q$}]
\addplot[m41] coordinates {(4,0.8994) (6,0.9057) (8,0.8987) (9,0.9153) (10,0.914) (12,0.9107) (14,0.9155)};
\addplot[m4o] coordinates {(4,0.9131) (5,0.9122) (6,0.8918) (7,0.9092) (8,0.917) (9,0.9162) (10,0.9206) (11,0.9164) (12,0.91) (14,0.9199)};
\addplot[mmini] coordinates {(4,0.8958) (5,0.9034) (6,0.8894) (7,0.875) (8,0.8424) (9,0.8299) (10,0.8511) (11,0.8206) (12,0.8161) (14,0.7814)};
\addplot[sel] coordinates {(9,0.9153) (10,0.9206) (5,0.9034)};
\nextgroupplot[ title={(b) \# calls ($\times10^3$)}, ylabel={\# calls ($\times10^3$)}]
\addplot[m41] coordinates {(4,5.992) (6,3.223) (8,2.233) (9,2.01) (10,1.805) (12,1.59) (14,1.534)};
\addplot[m4o] coordinates {(4,5.792) (5,4.026) (6,3.173) (7,2.655) (8,2.236) (9,2.007) (10,1.79) (11,1.683) (12,1.613) (14,1.536)};
\addplot[mmini] coordinates {(4,5.721) (5,4.019) (6,3.175) (7,2.666) (8,2.253) (9,2.154) (10,1.935) (11,1.965) (12,1.907) (14,2.005)};
\addplot[sel] coordinates {(9,2.01) (10,1.79) (5,4.019)};
\nextgroupplot[ title={(c) Tokens (M)}, ylabel={Tokens (M)}]
\addplot[m41] coordinates {(4,4.2776) (6,2.8897) (8,2.3187) (9,2.2046) (10,2.0617) (12,1.8779) (14,1.8322)};
\addplot[m4o] coordinates {(4,4.7384) (5,3.7005) (6,3.1701) (7,2.8541) (8,2.5742) (9,2.4234) (10,2.2413) (11,2.165) (12,2.0934) (14,2.0536)};
\addplot[mmini] coordinates {(4,3.9002) (5,3.0875) (6,2.6443) (7,2.3788) (8,2.1395) (9,2.1669) (10,2.0307) (11,2.1524) (12,2.1428) (14,2.4192)};
\addplot[sel] coordinates {(9,2.2046) (10,2.2413) (5,3.0875)};
\nextgroupplot[ymode=log, log basis y=10, title={(d) Cost (\$)}, ylabel={Cost (\$)}]
\addplot[m41] coordinates {(4,15.7309) (6,10.2211) (8,7.9526) (9,7.4932) (10,6.9166) (12,6.1778) (14,5.9569)};
\addplot[m4o] coordinates {(4,25.0684) (5,18.9778) (6,15.8722) (7,13.9905) (8,12.4087) (9,11.5001) (10,10.4911) (11,10.0392) (12,9.712) (14,9.571)};
\addplot[mmini] coordinates {(4,1.0164) (5,0.7712) (6,0.6332) (7,0.5439) (8,0.4746) (9,0.47) (10,0.4343) (11,0.4515) (12,0.4467) (14,0.4932)};
\addplot[sel] coordinates {(9,7.4932) (10,10.4911) (5,0.7712)};
\end{groupplot}
\end{tikzpicture}
\end{minipage}}\hfill
\fbox{\begin{minipage}[t]{0.480\textwidth}\centering
{\small\textbf{\em MovieLens}}\\[4pt]
\begin{tikzpicture}
\begin{groupplot}[group style={group size=2 by 2, horizontal sep=1.55cm, vertical sep=1.75cm}, sweep]
\nextgroupplot[ymin=0.84, ymax=0.90, ytick={0.84,0.86,...,0.90}, title={(a) $\mathrm{nDCG}_U$}, ylabel={$\mathrm{nDCG}_U$}]
\addplot[m41] coordinates {(4,0.8728) (5,0.8734) (6,0.8715) (7,0.8716) (8,0.8761) (10,0.8776) (11,0.8773) (12,0.8716) (13,0.8723) (14,0.8761)};
\addplot[m4o] coordinates {(4,0.8764) (5,0.8777) (6,0.8767) (7,0.8746) (8,0.8721) (9,0.8748) (10,0.8788) (11,0.872) (12,0.8721) (13,0.8757) (14,0.874)};
\addplot[mmini] coordinates {(4,0.8774) (5,0.8752) (6,0.8801) (7,0.876) (8,0.8758) (9,0.879) (10,0.8782) (11,0.8725) (12,0.8738) (13,0.8739) (14,0.8755)};
\addplot[sel] coordinates {(10,0.878) (10,0.879) (10,0.878)};
\nextgroupplot[ title={(b) \# calls ($\times10^3$)}, ylabel={\# calls ($\times10^3$)}]
\addplot[m41] coordinates {(4,6.281) (5,4.368) (6,3.256) (7,2.729) (8,2.373) (10,1.893) (11,1.815) (12,1.756) (13,1.728) (14,1.699)};
\addplot[m4o] coordinates {(4,5.818) (5,4.099) (6,3.094) (7,2.571) (8,2.219) (9,1.95) (10,1.832) (11,1.775) (12,1.74) (13,1.709) (14,1.677)};
\addplot[mmini] coordinates {(4,5.802) (5,4.004) (6,2.953) (7,2.454) (8,2.171) (9,1.923) (10,1.784) (11,1.756) (12,1.727) (13,1.703) (14,1.68)};
\addplot[sel] coordinates {(10,1.893) (10,1.832) (10,1.784)};
\nextgroupplot[ title={(c) Tokens (M)}, ylabel={Tokens (M)}]
\addplot[m41] coordinates {(4,2.4668) (5,1.7975) (6,1.4) (7,1.2103) (8,1.0803) (10,0.8965) (11,0.8636) (12,0.8385) (13,0.8299) (14,0.8125)};
\addplot[m4o] coordinates {(4,2.2987) (5,1.6989) (6,1.3383) (7,1.1477) (8,1.0174) (9,0.9127) (10,0.8688) (11,0.8474) (12,0.8306) (13,0.8192) (14,0.8049)};
\addplot[mmini] coordinates {(4,2.3055) (5,1.6732) (6,1.2905) (7,1.1055) (8,1.0033) (9,0.9078) (10,0.8584) (11,0.8463) (12,0.8316) (13,0.8256) (14,0.8104)};
\addplot[sel] coordinates {(10,0.9) (10,0.87) (10,0.86)};
\nextgroupplot[ymode=log, log basis y=10, title={(d) Cost (\$)}, ylabel={Cost (\$)}]
\addplot[m41] coordinates {(4,5.4816) (5,4.0575) (6,3.205) (7,2.7978) (8,2.5149) (10,2.1116) (11,2.0368) (12,1.9796) (13,1.9627) (14,1.9196)};
\addplot[m4o] coordinates {(4,6.4761) (5,4.877) (6,3.8982) (7,3.3827) (8,3.0234) (9,2.733) (10,2.6122) (11,2.5554) (12,2.5032) (13,2.4725) (14,2.4292)};
\addplot[mmini] coordinates {(4,0.3957) (5,0.2941) (6,0.2316) (7,0.2008) (8,0.1843) (9,0.1682) (10,0.1601) (11,0.1579) (12,0.1552) (13,0.1543) (14,0.1511)};
\addplot[sel] coordinates {(10,2.11) (10,2.61) (10,0.16)};
\end{groupplot}
\end{tikzpicture}
\end{minipage}}
\caption{RQ4: batch size~$m$ vs.\ quality, calls, tokens, and cost (\$). One line per model; stars mark selected~$m$ (star--line gap at the same~$m$ is run variation). Contiguous Round~1, diagonals then Pack, $\theta{=}1.0$; recursive $R_0$ except {\em DL19} Round-2 windows.}
\label{fig:m-sweep-datasets}
\label{fig:rq4-m-sweep}
\label{fig:rq4-batchsize}
\end{figure*}

We next ask how the listwise batch size~$m$ affects quality and cost across models and datasets, and how we select~$m$ for the RQ3 runs.
Figure~\ref{fig:m-sweep-datasets} sweeps $m$ for GPT-4.1, GPT-4o, and GPT-4o-mini on {\em NBA}, {\em Population}, {\em DL19}, and {\em MovieLens} (contiguous Round~1, diagonals then Pack, $\theta{=}1.0$), recording quality, calls, tokens, and USD cost; stars mark each model's selected~$m$.

\begin{center}
\captionof{table}{\small RQ3: TBS across LLMs at selected~$m$. Bold is best per row.}
\label{tab:rq3-models}
\label{tab:rq3-llms}
\scriptsize
\setlength{\tabcolsep}{3.2pt}
\renewcommand{\arraystretch}{1.05}
\begin{tabularx}{\columnwidth}{@{}l l >{\raggedleft\arraybackslash}X >{\raggedleft\arraybackslash}X >{\raggedleft\arraybackslash}X@{}}
\toprule
\textbf{Dataset} & \textbf{Metric} & \textbf{GPT-4.1} & \textbf{GPT-4o} & \textbf{GPT-4o-mini} \\
\midrule
\multirow{4}{*}{\em NBA}
  & $\tau$ $\uparrow$     & 0.817 & \textbf{0.876} & 0.208 \\
  & \# calls $\downarrow$   & 110 & \textbf{97} & 306 \\
  & Tokens (M) $\downarrow$ & \textbf{0.06} & \textbf{0.06} & 0.10 \\
  & Cost (\$) $\downarrow$  & 0.33 & 0.43 & \textbf{0.04} \\
\midrule
\multirow{4}{*}{\em Population}
  & $\tau$ $\uparrow$     & \textbf{0.972} & 0.932 & 0.952 \\
  & \# calls $\downarrow$   & \textbf{87} & 101 & 202 \\
  & Tokens (M) $\downarrow$ & \textbf{0.04} & 0.05 & 0.08 \\
  & Cost (\$) $\downarrow$  & 0.18 & 0.36 & \textbf{0.03} \\
\midrule
\multirow{4}{*}{\em DL19}
  & $\mathrm{nDCG}_Q$ $\uparrow$ & 0.905 & \textbf{0.921} & 0.900 \\
  & \# calls $\downarrow$   & 2{,}105 & \textbf{1{,}790} & 4{,}522 \\
  & Tokens (M) $\downarrow$ & 2.27 & \textbf{2.24} & 3.40 \\
  & Cost (\$) $\downarrow$  & 7.67 & 10.49 & \textbf{0.85} \\
\midrule
\multirow{4}{*}{\em SciFact}
  & $\mathrm{nDCG}_C$ $\uparrow$ & 0.864 & 0.839 & \textbf{0.871} \\
  & \# calls $\downarrow$   & \textbf{2{,}032} & 2{,}102 & 7{,}392 \\
  & Tokens (M) $\downarrow$ & \textbf{5.91} & 6.46 & 11.48 \\
  & Cost (\$) $\downarrow$  & 15.60 & 23.77 & \textbf{2.42} \\
\midrule
\multirow{4}{*}{\em MovieLens}
  & $\mathrm{nDCG}_U$ $\uparrow$ & 0.878 & \textbf{0.879} & 0.878 \\
  & \# calls $\downarrow$   & 1{,}893 & 1{,}832 & \textbf{1{,}784} \\
  & Tokens (M) $\downarrow$ & 0.90 & 0.87 & \textbf{0.86} \\
  & Cost (\$) $\downarrow$  & 2.11 & 2.61 & \textbf{0.16} \\
\bottomrule
\end{tabularx}
\end{center}

\spara{Cost vs.\ quality.}
Because each listwise call orders up to~$\binom{m}{2}$ pairs, increasing~$m$ shortens the three-round schedule and cuts calls, tokens, and dollar cost, with diminishing returns once~$m$ is already large.
Call counts for GPT-4.1, GPT-4o, and GPT-4o-mini nearly coincide at each~$m$, so the dollar gaps between models are driven mainly by token pricing rather than by how many batches \textsf{TBS} issues.
Quality, however, depends on whether the model can still produce a reliable total order over a longer list.
GPT-4.1 and GPT-4o remain essentially quality-flat across the sweep: larger batches reduce cost without a clear quality penalty, so the knee/endpoint rule of~\S\ref{sec:method:batchsize} selects large~$m$ ($9$--$12$) to harvest the call savings.
GPT-4o-mini does not share that robustness on total-order and passage ranking.
On {\em NBA}, mini already sits at $\tau{\approx}0.21$ for $m{=}4$ and collapses near zero for $m{\ge}6$; on {\em Population}, quality falls from ${\approx}0.97$ at small~$m$ to below~$0.6$ as~$m$ grows; on {\em DL19}, $\mathrm{nDCG}_Q$ drops once~$m$ exceeds~$5$--$6$.
In these settings a large~$m$ would buy fewer calls at the expense of a broken ranking, so the selector keeps mini at small batches ($m{=}4/6/5$ on {\em NBA}/{\em Population}/{\em DL19}).
{\em MovieLens} is the exception: preference ranking over short movie descriptors stays stable for all three models, and the selector picks $m{=}10$ for each.

\spara{Takeaway.}
When quality stays flat under longer lists, as with GPT-4.1 and GPT-4o, a larger~$m$ mainly reduces calls, tokens, and dollar cost.
GPT-4o-mini instead needs a small~$m$ on factual total-order and passage ranking, where longer listwise prompts degrade the weaker comparator.

\subsection{RQ5: Execution and Batching Strategies}
\label{sec:exp:rq5}
\label{sec:exp:batch-selection}
\label{sec:exp:passage-ablation}

Table~\ref{tab:rq5-r1} compares the four Round~1 membership presets of~\S\ref{sec:method:batchmembership} under GPT-4o ($m{=}10$, recursive $R_0$, diagonals then Pack, $\theta{=}1.0$).
Contiguous and random differ by only $0.002$--$0.024$ in quality across datasets, so the choice between them barely matters.
On {\em NBA} and {\em Population}, where keys are bare names with little text for clustering, both similarity presets trail contiguous on~$\tau$ while saving few calls.
On passage and preference tasks the text is richer, yet similarity still does not dominate: contiguous remains cheapest or near-tied on {\em DL19} and {\em MovieLens}, and only {\em SciFact} favors lexical similarity on quality.

\begin{center}
\captionof{table}{\small RQ5: Round~1 membership (GPT-4o, $m{=}10$). Bold is best per row.}
\label{tab:rq5-r1}
\scriptsize
\setlength{\tabcolsep}{2.2pt}
\renewcommand{\arraystretch}{0.90}
\begin{tabular}{@{}llrrrr@{}}
\toprule
\textbf{Dataset} & \textbf{Metric} & \textbf{Contig.} & \textbf{Random} & \textbf{Sim.} & \textbf{Sim.-emb.} \\
\midrule
\multirow{4}{*}{\em NBA}
 & $\tau$ $\uparrow$        & \textbf{0.876} & 0.866 & 0.849 & 0.832 \\
 & \# calls $\downarrow$      & 97 & 96 & 93 & \textbf{89} \\
 & Tokens (M) $\downarrow$    & 0.06 & 0.06 & 0.06 & \textbf{0.05} \\
 & Cost (\$) $\downarrow$     & 0.43 & 0.41 & 0.41 & \textbf{0.39} \\
\midrule
\multirow{4}{*}{\em Population}
 & $\tau$ $\uparrow$        & \textbf{0.932} & 0.908 & 0.861 & 0.892 \\
 & \# calls $\downarrow$      & 101 & \textbf{95} & \textbf{95} & 96 \\
 & Tokens (M) $\downarrow$    & 0.05 & 0.05 & 0.05 & 0.05 \\
 & Cost (\$) $\downarrow$     & 0.36 & 0.35 & \textbf{0.34} & \textbf{0.34} \\
\midrule
\multirow{4}{*}{\em DL19}
 & $\mathrm{nDCG}_Q$ $\uparrow$ & 0.921 & \textbf{0.923} & 0.895 & 0.897 \\
 & \# calls $\downarrow$      & \textbf{1{,}790} & 1{,}840 & 1{,}824 & 1{,}806 \\
 & Tokens (M) $\downarrow$    & \textbf{2.24} & 2.30 & 2.27 & 2.26 \\
 & Cost (\$) $\downarrow$     & \textbf{10.49} & 10.91 & 10.81 & 10.69 \\
\midrule
\multirow{4}{*}{\em SciFact}
 & $\mathrm{nDCG}_C$ $\uparrow$ & 0.839 & 0.861 & \textbf{0.862} & 0.838 \\
 & \# calls $\downarrow$      & 2{,}102 & 2{,}110 & 2{,}107 & \textbf{2{,}080} \\
 & Tokens (M) $\downarrow$    & 6.46 & 6.43 & 6.45 & \textbf{6.38} \\
 & Cost (\$) $\downarrow$     & 23.77 & 23.73 & 23.71 & \textbf{23.48} \\
\midrule
\multirow{4}{*}{\em MovieLens}
 & $\mathrm{nDCG}_U$ $\uparrow$ & \textbf{0.879} & 0.877 & 0.873 & 0.873 \\
 & \# calls $\downarrow$      & \textbf{1{,}832} & 1{,}855 & 1{,}854 & 1{,}841 \\
 & Tokens (M) $\downarrow$    & \textbf{0.87} & 0.88 & 0.88 & 0.88 \\
 & Cost (\$) $\downarrow$     & \textbf{2.61} & 2.66 & 2.66 & 2.64 \\
\bottomrule
\end{tabular}
\end{center}

\enlargethispage{5\baselineskip}
\spara{Takeaway.}
Round~1 membership changes quality more than call count, and contiguous or random is a stable choice on most datasets.
On {\em NBA}/{\em Population}, contiguous leads ($\tau{=}0.876$/$0.932$) and similarity trails by $0.027$--$0.071$ with little call savings, consistent with bare-name keys offering weak clustering signal.
On {\em DL19}/{\em MovieLens}, contiguous is cheapest or tied (random ties {\em DL19} at $0.923$); similarity trails {\em DL19} by ${\approx}0.025$ and stays within $0.006$ on {\em MovieLens}.
{\em SciFact} is the exception: lexical similarity ($0.862$) matches random and beats contiguous ($0.839$), likely because shared scientific terms put related abstracts in the same Round~1 column and make listwise comparisons easier, whereas embedding clusters are topical and need not align with claim-level support or refute; embedding similarity is cheapest overall but does not improve quality.

\clearpage

\end{document}